\documentclass[aps,prd,superscriptaddress,twoside,twocolumn,nofootinbib,showpacs,floatfix]{revtex4-1}
\usepackage{amsmath,amssymb}
\usepackage{graphicx}
\usepackage{subfigure}
\pdfpagewidth=\paperwidth
\pdfpageheight=\paperheight
\allowdisplaybreaks
\newcommand*{\slashed}[1]{{#1\!\!\!/}}
\newcommand*{\hc}{\text{H.\,c.}}

\begin{document}

\title{\boldmath Nucleon resonances in $\gamma p \to K^{*+} \Lambda$}

\author{A. C. Wang}
\affiliation{School of Physical Sciences, University of Chinese Academy of Sciences, Beijing 101408, China}

\author{W. L. Wang}
\affiliation{School of Physics and Nuclear Energy Engineering, Beihang University, Beijing 100191, China}
\affiliation{Department of Physics and Astronomy, University of Georgia, Athens, Georgia 30602, USA}

\author{F. Huang}
\email[Corresponding author. Email: ]{huangfei@ucas.ac.cn}
\affiliation{School of Physical Sciences, University of Chinese Academy of Sciences, Beijing 101408, China}

\author{H. Haberzettl}
\affiliation{Institute for Nuclear Studies and Department of Physics,
The George Washington University, Washington, DC 20052, USA}

\author{K. Nakayama}
\affiliation{Department of Physics and Astronomy, University of Georgia, Athens, Georgia 30602, USA}

\date{\today}

\begin{abstract}
The high-precision cross-section data for the reaction $\gamma p \to
K^{*+}\Lambda$ reported by the CLAS Collaboration at the Thomas Jefferson
National Accelerator Facility have been analyzed based on an effective
Lagrangian approach in the tree-level approximation. Apart from the
$t$-channel $K$, $\kappa$, $K^*$ exchanges, the $s$-channel nucleon ($N$)
exchange, the $u$-channel $\Lambda$, $\Sigma$, $\Sigma^*(1385)$ exchanges,
and the generalized contact term, the contributions from the near-threshold
nucleon resonances in the $s$-channel are also taken into account in
constructing the reaction amplitude. It is found that, to achieve a
satisfactory description of the differential cross section data, at least two
nucleon resonances should be included. By including the $N(2060){5/2}^-$
resonance, which is responsible for the shape of the angular distribution
near the $K^*\Lambda$ threshold, and one of the $N(2000){5/2}^+$,
$N(2040){3/2}^+$, $N(2100){1/2}^+$, $N(2120){3/2}^-$ and $N(2190){7/2}^-$
resonances, one can describe the cross-section data quite well, with the
fitted resonance masses and widths compatible with those advocated by the
Particle Data Group. The resulted predictions of the beam, target, and recoil
asymmetries are found to be quite different from various fits, indicating the
necessity of the spin observable data for $\gamma p \to K^{*+}\Lambda$ to
further pin down the resonance contents and associated parameters in this
reaction.
\end{abstract}



\maketitle

\section{Introduction}   \label{Sec:intro}

The extraction of nucleon resonances ($N^*$'s) from experimental data and
understanding their nature are essential to get insight into the
non-perturbative regime of Quantum Chromodynamics (QCD). Our current knowledge
of most of the $N^*$'s is mainly coming from the analyses of $\pi N$ scattering
and $\pi$ photoproduction off the nucleon. One of the problem with this
situation is that the quark models \cite{Isgur:1978,Capstick:1986,Loring:2001}
predict the existence of many more resonances than found in these reactions.
This is known as the \textit{missing resonance problem} \cite{Koniuk:1980}. The
number of baryon resonances in the lattice QCD calculations
\cite{Edwards:2013,Engel:2013} are also increasing.

Some of the nucleon resonances are known to couple weakly to the $\pi N$
channel, escaping their detections in these reactions. This forces us to search
for those missing resonances in channels other than $\pi N$, where they couple
more strongly so that they can be better established. In the present work we
investigate the $K^*\Lambda$ photoproduction reaction in search for clear
evidence of resonances that may be revealed through their couplings to the
$K^*\Lambda$ channel. There are many attractive features  in studying this
reaction. First of all, resonances with sizable hidden $s\bar{s}$ content can
have a better chance to be revealed in this reaction than in $\pi$ production
reactions. Also, since the threshold of $K^*\Lambda$ is much higher than that
of $\pi N$, the $K^*\Lambda$ photoproduction off nucleon is more suited than
the $\pi$ production reactions for investigating the nucleon resonances in a
less-explored higher $N^*$ mass region. Another advantage of $K^*\Lambda$
photoproduction in studying $N^*$'s is that it acts as an ``isospin filter"
isolating the $N^*$'s with isospin $I=1/2$.

Experimentally, so far the available data for the reaction $\gamma p \to
K^{*+}\Lambda$ are all reported by the CLAS Collaboration at the Thomas
Jefferson National Accelerator Facility (JLab). The first preliminary total
cross section data for center-of-mass energy, $W$, from threshold up to 2.85
GeV were reported by Guo {\it et al.} in 2006 in a conference proceedings
\cite{Guo:2006}. Later, the preliminary differential cross section data for
this reaction from $W= 2.22$ GeV to $2.42$ GeV were reported by Hicks {\it et
al.} in 2011 in another conference proceedings \cite{Hicks:2011}. It was only
in 2013 that the first high-statistics cross section data for this reaction
were published by Tang {\it et al.} in Ref.~\cite{Tang:2013}, where the
measured differential cross sections and the extracted total cross sections are
presented from threshold up to $W\approx 2.85$ GeV. Also, a few preliminary
differential cross section data for the $\gamma n \to K^{*0} \Lambda$ reaction
have been reported by Mattione in a conference proceedings
\cite{Mattione:2014}.

The CLAS differential cross section data for $\gamma p \to K^{*+}\Lambda$
\cite{Tang:2013} show some structures near the $K^{*+}\Lambda$ threshold energy
which may indicate some possible contribution from nucleon resonance(s). In
fact, in this energy region, there are six resonances advocated in the most
recent Particle Data Group (PDG) review \cite{Patrignani:2016} that might
potentially contribute to this reaction, namely, $N(2000){5/2}^+$,
$N(2040){3/2}^+$, $N(2060){5/2}^-$, $N(2100){1/2}^+$, $N(2120){3/2}^-$, and
$N(2190){7/2}^-$. Among them, $N(2190){7/2}^-$ is rated as a four-star
resonance but with rather broad mass ($2100-2200$ MeV) and width ($300-700$
MeV); $N(2000){5/2}^+$, $N(2060){5/2}^-$ and $N(2120){3/2}^-$ are rated as
two-star resonances and, $N(2040){3/2}^+$ and $N(2100){1/2}^+$ as one-star
resonances. This means that the four-star $N(2190){7/2}^-$ resonance needs
further investigation to improve the accuracy of its parameters, and the other
five two- and one-star resonances need more information, especially from the
reaction channels (other than those cited in PDG) to which these resonances
couple more strongly, to improve the evidences of their existences and to
extract their parameters. The high-statistics cross section data for $\gamma
p\to K^{*+}\Lambda$ from the CLAS Collaboration \cite{Tang:2013} promote the
studies along this direction.

\begin{figure*}[tbp]
\includegraphics[width=0.65\textwidth]{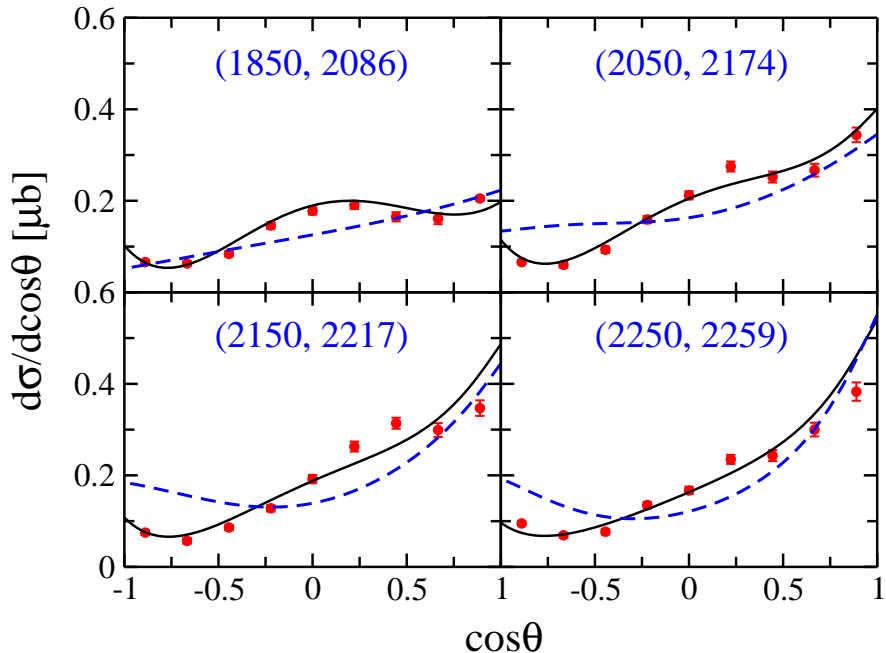}
\caption{(Color online) Status of theoretical description of the differential
cross sections for $\gamma p\to K^{*+}\Lambda$ at selected energies in the near
threshold region. The numbers in parentheses denote the photon laboratory
incident energy (left number) and the total center-of-mass energy of the system
(right number). The blue dashed lines represent the results from
Ref.~\cite{Kim:2014}, and the black solid lines denote the results from model I
of our present work which will be discussed later. The scattered symbols are
the most recent data from CLAS Collaboration \cite{Tang:2013}.}
\label{Fig:status}
\end{figure*}

Theoretically, several works based on effective Lagrangian approaches have
already been devoted to the study of $K^*\Lambda$ photoproduction reaction
\cite{Oh-1:2006,Oh-2:2006,Ozaki:2010,Kim:2011,Kim:2014,Yu:2016}. In 2006, Oh
and Kim have investigated the non-resonant contributions for $\gamma N\to
K^*\Lambda$ within an isobar model, and they found that the $t$-channel
$K$-exchange, which causes a sharp raise of the differential cross sections at
forward-scattering angles, dominates this reaction process \cite{Oh-1:2006}.
Further in late 2006, they have examined the contribution of scalar meson
$\kappa$, and concluded that the $t$-channel $\kappa$-exchange in $\gamma p \to
K^{*+}\Lambda$ is rather small \cite{Oh-2:2006}. In 2010, Ozaki {\it et al.}
have studied the $\gamma p \to K^{*+}\Lambda$ reaction in a Regge model
\cite{Ozaki:2010}. They have obtained the total cross sections compatible with
CLAS's preliminary data \cite{Guo:2006}, and found that the contributions from
the $K^*$ trajectory and reggeized contact term are much bigger than those in
the isobar model of Ref.~\cite{Oh-1:2006}. (However, we point out here
that the Regge model of Ref.~\cite{Guidal:1997} used in Ref.~\cite{Ozaki:2010}
is based on incorrect dynamical assumptions, as shown in Ref.~\cite{Haberzettl:2015}.)
In 2011, Kim {\it et al.} \cite{Kim:2011} have included the contributions from
the resonances $N(2080){3/2}^-$ and $N(2200){5/2}^-$ based on the theoretical
models of Refs.~\cite{Oh-1:2006,Oh-2:2006} in order to describe the preliminary
differential cross section data from CLAS \cite{Hicks:2011}. They have found
that the non-resonant contributions dominate the $K^*\Lambda$ photoproduction
reaction, while the resonance $N(2080){3/2}^-$ plays a crucial role in
explaining the enhancement of the near-threshold production rate and the
contribution from $N(2200){5/2}^-$ is rather small. When the first
high-statistics cross section data from CLAS was published in 2013
\cite{Tang:2013}, it was found that all the above-mentioned theoretical
calculations \cite{Oh-1:2006,Oh-2:2006,Ozaki:2010,Kim:2011} significantly
underestimate the cross sections in the range of (laboratory) photon energy of
2.1 GeV $< E_\gamma <$ 3.1 GeV. Then in 2014, Kim {\it et al.} have
re-investigated \cite{Kim:2014} the $K^*\Lambda$ photoproduction reaction to
accommodate the most recent CLAS data \cite{Tang:2013}. They have considered
four nucleon resonances, namely $N(2000){5/2}^+$, $N(2060){5/2}^-$,
$N(2120){3/2}^-$ and $N(2190){7/2}^-$, in addition to the non-resonant
contributions as included in Ref.~\cite{Kim:2011}, and found that apart from
the significant contributions from the $t$-channel $K$ and $\kappa$ exchanges,
the $s$-channel nucleon resonances $N(2120){3/2}^-$ and $N(2190){7/2}^-$ play
very important roles in reproducing the experimental cross section data. The
contribution from the resonance $N(2060){5/2}^-$ was found to be small but
noticeable, while that from $N(2000){5/2}^+$ was found to be almost negligible.
In Ref.~\cite{Yu:2016}, the total cross sections and the differential cross
sections at three selected energies for $\gamma N \to K^* \Lambda$ are
investigated within a Regge approach. (The dynamical assumptions~\cite{Guidal:1997} of this
Regge analysis are also marred by incomplete dynamical assumptions~\cite{Haberzettl:2015}.)
It is found that the $K$ and $K^*$ trajectories dominate the the process of
$K^{*+}\Lambda$ photoproduction. The preliminary differential cross section
data for $\gamma n \to K^{*0} \Lambda$ \cite{Mattione:2014} have been also
analyzed recently by Wang and He \cite{Wang:2016} in an effective Lagrangian
approach.

The work of Ref.~\cite{Kim:2014} presents so far the only detailed theoretical
analysis of the most recent high-statistics differential cross section data for
$\gamma p\to K^{*+}\Lambda$ reported by the CLAS Collaboration
\cite{Tang:2013}. It describes the total cross section data quite well in the
photon energy region of $E_\gamma < 3.5$ GeV, and the differential cross
section data have also been qualitatively described. Nevertheless, there is
still some room for improvement in their results for the differential cross
sections, especially, near the $K^{*+}\Lambda$ threshold, where the nucleon
resonances are relevant. Figure~\ref{Fig:status} illustrates this issue; there,
a comparison of the differential cross sections from the theoretical
calculation of Ref.~\cite{Kim:2014} (blue dashed lines) with the most recent
CLAS data \cite{Tang:2013} (scattered symbols) at some selected energies in the
near-threshold region is shown. The numbers in parentheses denote the photon
laboratory incident energy, $E_\gamma$, (left number) and the total
center-of-mass energy of the system, $W$, (right number). The black solid lines
represent the results from model I of our present work which will be discussed
in detail in Sec.~\ref{Sec:results}. It is clearly seen from
Fig.~\ref{Fig:status} that there is still some room for improvement in the
differential cross section results of Ref.~\cite{Kim:2014}. We mention that in 
Ref.~\cite{Kim:2014} the resonance parameters of $N(2000){5/2}^+$ and $N(2060){5/2}^-$ are taken from
Ref.~\cite{Anisovich:2013}, the parameters of $N(2190){7/2}^-$ are taken from Ref.~\cite{Capstick:1992}, and the parameters of $N(2120){3/2}^-$ are determined by a fit to the experimental data. 

In this work, we investigate the $\gamma p \to K^{*+}\Lambda$ reaction based on
an effective Lagrangian approach in the tree-level approximation. We expect that 
a better description of the data for this reaction will allow for a
more reliable extraction of the resonance content and their associated parameters.
One of the major differences of our theoretical model compared with that of
Refs.~\cite{Kim:2011,Kim:2014} is that in the latter a common form factor is
introduced in the reaction amplitudes in order to preserve gauge invariance,
while in our work, following
Refs.~\cite{Haberzettl:1997,Haberzettl:2006,Huang:2012,Huang:2013}, a
generalized contact current -- that accounts effectively for the interaction
current arising from the unknown parts of the underlying microscopic model --
is introduced in such a way that the full photoproduction amplitude satisfies
the generalized Ward-Takahashi-Identity (WTI) and thus it is fully gauge
invariant. As a consequence, our model is free from such an artificial
constraint as the use of a common form factor. Moreover, and most relevantly,
we adopt a rather different strategy in choosing the nucleon resonances to be
considered in our model. Instead of including all of them,
we introduce the nucleon resonances in the present work as few as possible with
the resonance parameters being adjusted to reproduce the data. We find that
apart from the $t$-channel $K$, $\kappa$, $K^*$ exchanges, the $s$-channel
nucleon ($N$) exchange, the $u$-channel $\Lambda$, $\Sigma$, $\Sigma^*(1385)$
exchanges, and the generalized contact current, at least two nucleon resonances
near the $K^*\Lambda$ threshold should be included in the $s$-channel in order
to obtain a satisfactory description of the CLAS high-statistics differential
cross section data. By including the $N(2060){5/2}^-$ resonance, which, as we
shall show later, is responsible for the shape of the angular distribution near
the $K^*\Lambda$ threshold, and one of the $N(2000){5/2}^+$, $N(2040){3/2}^+$,
$N(2100){1/2}^+$, $N(2120){3/2}^-$ and $N(2190){7/2}^-$ resonances, we get five
fits with roughly the similar fit qualities. The resulting differential and
total cross sections are both in very good agreement overall with the most
recent CLAS data \cite{Tang:2013}. In particular, the angular dependence of the
differential cross sections near the $K^{*+}\Lambda$ threshold is now, for the
first time, described quite well. The fitted resonance masses and widths are
compatible with those advocated by the PDG \cite{Patrignani:2016}. The
non-resonant terms, dominated by the $t$-channel $K$ exchange, are found to
have very significant contributions. The predictions for the photon beam
asymmetry, target nucleon asymmetry, and recoil $\Lambda$ asymmetry are also
given; they are found to be more sensitive to the details of the model than the
cross sections, indicating the necessity of data on these spin observables to
further constrain the resonance contents and their parameters in this reaction.

Of course, a more complete analysis and extraction of nucleon resonances requires a coupled channels approach \cite{Anisovich:2012,Ronchen:2013,Shklyar:2013,Kamano:2015,Ramirez:2016}, so far developed mostly for pseudo-scalar meson production reactions. In this approach, the unitarity and analyticity of the reaction amplitude can be maintained and the search of poles (associated with the resonances) in the complex energy plane can be performed. This is beyond the scope of the present work which may be considered as a first step toward developing such a more complete model.

The present paper is organized as follows. In Sec.~\ref{Sec:formalism}, we briefly introduce the framework of our theoretical model. There, the strategy for imposing gauge invariance of the photoproduction amplitude according to the generalized WTI, the specific forms of the effective interaction Lagrangians, the resonance propagators and the phenomenological form factors are explicitly presented. In Sec.~\ref{Sec:results}, the results of our model calculations are shown, including a comparison of our calculated cross sections with the most recent high-statistics CLAS data, an analysis of the $\gamma p \to K^{*+}\Lambda$ reaction dynamics, and a discussion of the resulting resonance contents and associated parameters. Our predicted beam, target, and recoil asymmetries in $\gamma p \to K^{*+}\Lambda$ are also shown and discussed in this section. Finally a brief summary and conclusions are given in Sec.~\ref{sec:summary}.

\section{Formalism}  \label{Sec:formalism}

\begin{figure}[tbp]
\centering
{\vglue 0.15cm}
\subfigure[~$s$ channel]{
\includegraphics[width=0.45\columnwidth]{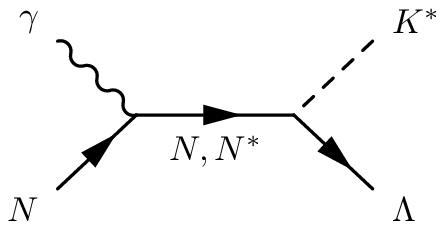}}  {\hglue 0.4cm}
\subfigure[~$t$ channel]{
\includegraphics[width=0.45\columnwidth]{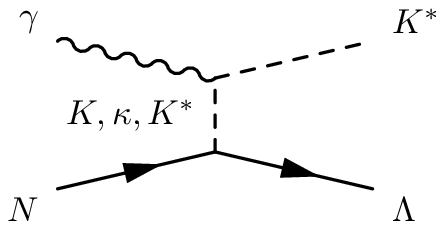}} \\[6pt]
\subfigure[~$u$ channel]{
\includegraphics[width=0.45\columnwidth]{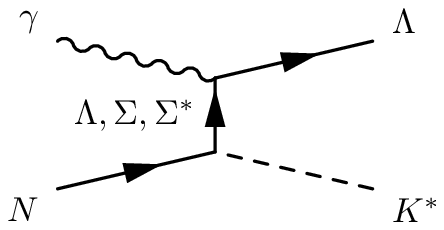}} {\hglue 0.4cm}
\subfigure[~Interaction current]{
\includegraphics[width=0.45\columnwidth]{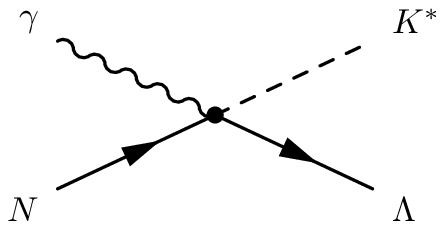}}
\caption{Generic structure of the $K^*$ photoproduction amplitude for $\gamma N\to K^{*}\Lambda$. Time proceeds from left to right.}
\label{FIG:feymans}
\end{figure}

Following a full field theoretical approach of Refs.~\cite{Haberzettl:1997,Haberzettl:2006,Huang:2012,Huang:2013}, the full reaction amplitude for $\gamma N \to K^* \Lambda$ can be expressed as
\begin{eqnarray}
M^{\nu\mu} = M^{\nu\mu}_s + M^{\nu\mu}_t + M^{\nu\mu}_u + M^{\nu\mu}_{\rm int},  \label{eq:amplitude}
\end{eqnarray}
with $\nu$ and $\mu$ being the Lorentz indices of vector meson $K^*$ and photon $\gamma$, respectively. The first three terms $M^{\nu\mu}_s$, $M^{\nu\mu}_t$, and $M^{\nu\mu}_u$ stand for the $s$-, $t$-, and $u$-channel pole diagrams, respectively, with $s$, $t$, and $u$ being the Mandelstam variables of the internally exchanged particles. They arise from the photon attaching to the external particles  in the underlying $\Lambda NK^*$ interaction vertex. The last term, $M^{\nu\mu}_{\rm int}$, stands for the interaction current which arises from the photon attaching to the internal structure of the $\Lambda NK^*$ interaction vertex. All four terms in Eq.~(\ref{eq:amplitude}) are diagrammatically depicted in Fig.~\ref{FIG:feymans}.

In the present work, the following contributions, as shown in Fig.~\ref{FIG:feymans}, are considered in constructing the $s$-, $t$-, and $u$-channel amplitudes: (a) $N$ and $N^*$'s exchanges in the $s$-channel, (b) $K$, $\kappa$, and $K^*$ meson exchanges in the $t$-channel, and (c) $\Lambda$, $\Sigma$, and $\Sigma^*(1385)$ hyperon exchanges in the $u$-channel. 
The exchanges of other hyperon resonances with higher masses in the $u$-channel are tested to have tiny contributions and thus are omitted in the present work in order to reduce the model parameters. Using an effective Lagrangian approach, one can, in principle, obtain explicit expressions for these amplitudes. However, the exact calculation of the interaction current $M^{\nu\mu}_{\rm int}$ is impractical, as it obeys a highly non-linear equation and contains diagrams with very complicated interaction dynamics. Furthermore, the introduction of phenomenological form factors makes it impossible to calculate the interaction current exactly even in principle. Following Refs.~\cite{Haberzettl:1997,Haberzettl:2006,Huang:2012}, we model the interaction current by a generalized contact current,
\begin{eqnarray}
M^{\nu\mu}_{\rm int} = \Gamma^\nu_{\Lambda N K^*}(q) C^\mu + M_{\rm KR}^{\nu\mu} f_t.  \label{eq:Mint}
\end{eqnarray}
Here $\nu$ and $\mu$ are Lorentz indices for $K^*$ and $\gamma$, respectively; $\Gamma^\nu_{\Lambda N K^*}(q)$ is the vertex function of $\Lambda N K^*$ coupling given by the Lagrangian of Eq.~(\ref{eq:L_LNKst}),
\begin{eqnarray}
\Gamma^\nu_{\Lambda N K^*}(q) = - i g_{\Lambda N K^*}\left[\gamma^{\nu}-i\frac{\kappa_{\Lambda N K^*}}{2M_N}\sigma^{\nu \alpha}q_{\alpha}\right]
\end{eqnarray}
with $q$ being the 4-momentum of the outgoing $K^*$ meson; $M_{\rm KR}^{\nu\mu}$ is the Kroll-Ruderman term given by the Lagrangian of Eq.~(\ref{eq:L_gLNKst}),
\begin{eqnarray}
M_{\rm KR}^{\nu \mu}= g_{\Lambda N K^*}\frac{\kappa_{\Lambda N K^*}}{2M_N}\sigma^{\nu\mu} Q_{K^*},
\end{eqnarray}
with $Q_{K^*}$ being the electric charge of $K^*$; $f_t$ is the phenomenological form factor attached to the amplitude of $t$-channel $K^*$-exchange, which is given in Eq.~(\ref{eq:ff_M}); $C^\mu$ is an auxiliary current, which is non-singular, introduced to ensure that the full photoproduction amplitude of Eq.~(\ref{eq:amplitude}) satisfies the generalized WTI and thus is fully gauge invariant. Following Refs.~\cite{Haberzettl:2006,Huang:2012}, we choose $C^\mu$ for $\gamma p \to K^{*+} \Lambda$ as
\begin{equation}
C^\mu =  - Q_{K^*} \frac{f_t-\hat{F}}{t-q^2}  (2q-k)^\mu - Q_N \frac{f_s-\hat{F}}{s-p^2} (2p+k)^\mu,
\end{equation}
with
\begin{equation} \label{eq:Fhat}
\hat{F} = 1 - \hat{h} \left(1 -  f_s\right) \left(1 - f_t\right).
\end{equation}
Here $p$, $q$, and $k$ are 4-momenta for incoming $N$, outgoing $K^*$ and incoming photon, respectively; $Q_{N\left(K^*\right)}$ is the electric charge of $N\left(K^*\right)$; $f_s$ is the phenomenological form factor for $s$-channel $N$-exchange.  
$\hat{h}$ is an arbitrary function, except that it should go to unity in the high-energy limit to prevent the ``violation of scaling behavior" \cite{Drell:1972}. For the sake of simplicity, in the present work it is taken to be $\hat{h}=1$.

In the rest of this section,  we present the effective Lagrangians, the resonance propagators and the phenomenological form factors employed in the present work.

\subsection{Effective Lagrangians} \label{Sec:Lagrangians}

The effective interaction Lagrangians used in the present work for the production amplitudes are given below. For further convenience, we define the operators
\begin{equation}
\Gamma^{(+)}=\gamma_5  \quad  \text{and} \quad  \Gamma^{(-)}=1,
\end{equation}
and the field-strength tensors
\begin{eqnarray}
{K^*}^{\mu\nu} &=& \partial^{\mu} {K^*}^{\nu} - \partial^{\nu} {K^*}^{\mu},  \\[6pt]
F^{\mu\nu} &=& \partial^{\mu}A^\nu-\partial^{\nu}A^\mu,
\end{eqnarray}
with ${K^{*\mu}}$ and $A^\mu$ denoting the $K^*$ vector-meson field and electromagnetic field, respectively.

The electromagnetic interaction Lagrangians required to calculate the non-resonant Feynman diagrams are
\begin{eqnarray}
{\cal L}_{NN\gamma} &=& -\,e \bar{N} \left[ \left( \hat{e} \gamma^\mu - \frac{ \hat{\kappa}_N} {2M_N}\sigma^{\mu \nu}\partial_\nu\right) A_\mu\right] N, \\[6pt]
{\cal L}_{\gamma K^* K^* } &=& -\,e \left({K^*}^{\nu} \times K^*_{\mu\nu}\right)_3 A^\mu, \\[6pt]
{\cal L}_{\gamma \kappa{K^*}} &=& e\frac{g_{\gamma \kappa{K^*}}}{2M_{K^*}}F^{\mu \nu}K^*_{\mu \nu}\kappa, \label{Lag:gkaKst}    \\[6pt]
{\cal L}_{\gamma K{K^*}} &=& e\frac{g_{\gamma K{K^*}}}{M_K}\varepsilon^{\alpha \mu \lambda \nu}\left(\partial_\alpha A_\mu\right)\left(\partial_\lambda K\right)K^*_\nu, \label{Lag:gKKst} \\[6pt]
{\cal L}_{\Lambda \Lambda \gamma} &=& e\frac{\kappa_\Lambda}{2M_N} \bar{\Lambda} \sigma^{\mu \nu}\left(\partial_{\nu}A_\mu\right)\Lambda,   \\[6pt]
{\cal L}_{\Sigma \Lambda \gamma} &=& e\frac{\kappa_{\Sigma \Lambda}}{2M_N}\bar{\Lambda}\sigma^{\mu \nu}\left(\partial_\nu A_\mu\right)\Sigma^0 + \hc,  \\[6pt]
{\cal L}_{\Sigma^* \Lambda \gamma} &=& ie\frac{g^{(1)}_{\Sigma^* \Lambda \gamma}}{2M_N}\bar{\Lambda}\gamma_\nu \gamma_5 F^{\mu \nu} \Sigma^{*0}_\mu \nonumber \\
&& -\,e\frac{g^{(2)}_{\Sigma^* \Lambda \gamma}}{\left(2M_N\right)^2} \left(\partial_\nu\bar{\Lambda}\right) \gamma_5 F^{\mu \nu}{\Sigma}^{*0}_\mu + \hc,
\end{eqnarray}
where $e$ is the elementary charge unit and $\hat{e}$ stands for the charge operator; $\hat{\kappa}_N = \kappa_p\left(1+\tau_3\right)/2 + \kappa_n\left(1-\tau_3\right)/2$, with the anomalous magnetic moments $\kappa_p=1.793$ and $\kappa_n=-1.913$; $\kappa_\Lambda=-0.613$ is the $\Lambda$ anomalous magnetic moment and $\kappa_{\Sigma\Lambda}=-1.61$ is the anomalous magnetic moment for $\Sigma^0\to \Lambda \gamma$ transition; $M_N$, $M_K$ and $M_{K^*}$ stand for the masses of $N$, $K$ and $K^*$, respectively; $\varepsilon^{\alpha \mu \lambda \nu}$ is the totally antisymmetric Levi-Civita tensor with $\varepsilon^{0123}=1$. The coupling constant $g_{\gamma\kappa K^*}=0.214$ is taken from Refs.~\cite{Kim:2011,Kim:2014}, determined by a vector-meson dominance model proposed by D. Black {\it et al.} \cite{BHS02}. The value of the electromagnetic coupling $g_{\gamma K K^*}$ is determined by fitting the radiative decay width of $K^*\to K\gamma$ given by the PDG \cite{Patrignani:2016}, which leads to $g_{\gamma K^\pm K^{*\pm}}=0.413$, with the sign inferred from $g_{\gamma \pi \rho}$ \cite{Garcilazo:1993} via the flavor SU(3) symmetry considerations in conjunction with the vector-meson dominance assumption. The electromagnetic couplings $g^{(1)}_{\Sigma^* \Lambda \gamma}$ and $g^{(2)}_{\Sigma^* \Lambda \gamma}$ should, in principle, be fixed by the helicity amplitudes of the transition reaction $\Sigma^{*0} \to \Lambda \gamma$. Nevertheless, the latest PDG \cite{Patrignani:2016} is still devoid of such information, and thus we treat the coupling $g^{(1)}_{\Sigma^* \Lambda \gamma}$ as a fit parameter and  let the coupling $g^{(2)}_{\Sigma^* \Lambda \gamma}$ be determined by the PDG value of the partial decay width, $\Gamma_{\Sigma^{*0} \to \Lambda \gamma}=0.45$ MeV \cite{Patrignani:2016}.

The resonance-nucleon-photon transition Lagrangians are
\begin{eqnarray}
{\cal L}_{RN\gamma}^{1/2\pm} &=& e\frac{g_{RN\gamma}^{(1)}}{2M_N}\bar{R} \Gamma^{(\mp)}\sigma_{\mu\nu} \left(\partial^\nu A^\mu \right) N  + \hc, \\[6pt]
{\cal L}_{RN\gamma}^{3/2\pm} &=& -\, ie\frac{g_{RN\gamma}^{(1)}}{2M_N}\bar{R}_\mu \gamma_\nu \Gamma^{(\pm)}F^{\mu\nu}N \nonumber \\
&&+\, e\frac{g_{RN\gamma}^{(2)}}{\left(2M_N\right)^2}\bar{R}_\mu \Gamma^{(\pm)}F^{\mu \nu}\partial_\nu N + \hc, \\[6pt]
{\cal L}_{RN\gamma}^{5/2\pm} & = & e\frac{g_{RN\gamma}^{(1)}}{\left(2M_N\right)^2}\bar{R}_{\mu \alpha}\gamma_\nu \Gamma^{(\mp)}\left(\partial^{\alpha} F^{\mu \nu}\right)N \nonumber \\
&& \pm\, ie\frac{g_{RN\gamma}^{(2)}}{\left(2M_N\right)^3}\bar{R}_{\mu \alpha} \Gamma^{(\mp)}\left(\partial^\alpha F^{\mu \nu}\right)\partial_\nu N \nonumber \\
&& + \, \hc,  \\[6pt]
{\cal L}_{RN\gamma}^{7/2\pm} &=&  ie\frac{g_{RN\gamma}^{(1)}}{\left(2M_N\right)^3}\bar{R}_{\mu \alpha \beta}\gamma_\nu \Gamma^{(\pm)}\left(\partial^{\alpha}\partial^{\beta} F^{\mu \nu}\right)N \nonumber \\
&&-\, e\frac{g_{RN\gamma}^{(2)}}{\left(2M_N\right)^4}\bar{R}_{\mu \alpha \beta} \Gamma^{(\pm)} \left(\partial^\alpha \partial^\beta F^{\mu \nu}\right) \partial_\nu N  \nonumber \\
&&  + \, \hc,
\end{eqnarray}
where $R$ designates the nucleon resonance, and the superscript of ${\cal L}_{RN\gamma}$ denotes the spin and parity of the resonance $R$. The coupling constants $g_{RN\gamma}^{(i)}$ $(i=1,2)$ are fit parameters.

The effective Lagrangians for meson-baryon interactions are
\begin{eqnarray}
{\cal L}_{\Lambda N {K^*}} &=&  -\, g_{\Lambda N {K^*}} \bar{\Lambda} \left[\left(\gamma^\mu-\frac{\kappa_{\Lambda N {K^*}}}{2M_N}\sigma^{\mu \nu}\partial_\nu\right)K^*_\mu\right] N \nonumber \\
&& +\, \hc,   \label{eq:L_LNKst}  \\[6pt]
{\cal L}_{\Lambda N\kappa} &=& -\, g_{\Lambda N\kappa} \bar{\Lambda} \kappa N + \hc, \\[6pt]
{\cal L}_{\Lambda NK} &=& -\, g_{\Lambda NK}\bar{\Lambda}\Gamma^{(+)} \left[\left(i\lambda + \frac{1-\lambda}{2M_N} \slashed{\partial}\right)K\right] N \nonumber \\
&& +\, \hc,  \label{eq:L_LNK}    \\[6pt]
{\cal L}_{\Sigma N {K^*}} &=& -\, g_{\Sigma N {K^*}}\bar{\Sigma}\left[\left(\gamma^\mu-\frac{\kappa_{\Sigma N {K^*}}}{2M_N}\sigma^{\mu \nu}\partial_\nu\right)K^*_\mu\right] N \nonumber \\
&& +\, \hc, \\[6pt]
{\cal L}_{\Sigma^* N{K^*}} &=& -\, i\frac{g_{\Sigma^* N K^*}^{(1)}}{2M_N}{\bar\Sigma}^*_\mu \gamma_\nu \gamma_5 {K^*}^{\mu \nu}N \nonumber \\
&& +\, \frac{g_{\Sigma^* N K^*}^{(2)}}{\left(2M_N\right)^2}{\bar\Sigma}^*_\mu \gamma_5 {K^*}^{\mu \nu}\partial_\nu N \nonumber \\
&& -\, \frac{g_{\Sigma^* N K^*}^{(3)}}{\left(2M_N\right)^2}{\bar\Sigma}^*_\mu \gamma_5\left(\partial_\nu {K^{*\mu \nu}}\right) N + \hc.
\end{eqnarray}
where the parameter $\lambda$ was introduced in ${\cal L}_{\Lambda NK}$ to interpolate between the pseudo-vector $(\lambda=0)$ and the pseudo-scalar $(\lambda=1)$ couplings. Unlike for the pion coupling, where the low-energy chiral perturbation theory calls for the pseudo-vector coupling over the pseudo-scalar coupling, for Kaons, the situation is much less clear. In fact, some authors have employed pseudo-scalar coupling \cite{Kim:2014}
and others have allowed for both types of couplings \cite{Nakayama:2006Xi}. On the other hand, it is a common practice to rely on SU(3) flavor symmetry for obtaining the effective Lagrangians when studying the Kaon-baryon systems, which implies a pseudo-vector $\Lambda NK$ coupling, since the pseudo-vector coupling is used in the $NN\pi$ vertex as demanded by chiral symmetry.  For example, Haidenbauer \textit{et al.} \cite{Haidenbauer:2013}, have obtained an excellent description of the hyperon-nucleon system in chiral effective field theory, i.e.,  with pseudo-vector $\Lambda NK$ coupling. In the present work, following Refs.~\cite{Haidenbauer:2013,Ronchen:2013}, $\lambda$ is set to be zero, i.e., we adopt the pure pseudo-vector type coupling. Although we shall not show any results with the pseudo-scalar coupling for the $\Lambda NK$ vertex in the present work, we just mention that we have tested this coupling type during the trial calculations and found that it leads to results that are worse than those obtained using the pseudo-vector coupling. The coupling constant $g_{\Lambda NK}=13.99$ is taken from Ref.~\cite{Ronchen:2013}, determined by the flavor SU(3) symmetry. The coupling constants $g_{\Lambda N {K^*}}$, $\kappa_{\Lambda N {K^*}}$, $g_{\Sigma N {K^*}}$, $\kappa_{\Sigma N {K^*}}$ and $g^{(1)}_{\Sigma^* N K^*}$ are also fixed by the flavor SU(3) symmetry \cite{Swart:1963,Ronchen:2013},
\begin{eqnarray}
g_{\Lambda N K^*} &=&  -\, \frac{1}{2\sqrt{3}} g_{NN\omega} - \frac{\sqrt{3}}{2} g_{NN\rho} = -6.21,  \label{g_LNKst} \\[6pt]
\kappa_{\Lambda N K^*} &=&  \frac{f_{\Lambda N K^*}}{g_{\Lambda N K^*}}  = -\frac{\sqrt{3}}{2} \frac{f_{NN\rho}}{g_{\Lambda N K^*}} = 2.76,  \label{kappa_LNKst}  \\[6pt]
g_{\Sigma N K^*} &=&  -\, \frac{1}{2} g_{NN\omega} + \frac{1}{2} g_{NN\rho} = -4.26,  \\[6pt]
\kappa_{\Sigma N K^*} &=&  \frac{f_{\Sigma N K^*}}{g_{\Sigma N K^*}}  = \frac{1}{2} \frac{f_{NN\rho}}{g_{\Sigma N K^*}} = -2.33, \\[6pt]
g^{(1)}_{\Sigma^* NK^*} &=& -\,\frac{1}{\sqrt{6}} g_{\Delta N\rho} = 15.96,
\end{eqnarray}
where the empirical values $g_{NN\rho}=3.25$, $g_{NN\omega}=11.76$, $\kappa_{NN\rho}=g_{NN\rho}/f_{NN\rho}=6.1$ and $g_{\Delta N\rho}=-39.10$ from Refs.~\cite{Huang:2012,Ronchen:2013} are quoted. As the $g^{(2)}$ and $g^{(3)}$ terms in the $\Delta N\rho$ interactions have never been seriously studied in literature, the corresponding couplings for the $\Sigma^* NK^*$ interactions, i.e. $g_{\Sigma^* N K^*}^{(2)}$ and $g_{\Sigma^* N K^*}^{(3)}$, cannot be determined via flavor SU(3) symmetry, and we ignore these two terms in the present work, following Refs.~\cite{Kim:2011,Kim:2014}. The coupling constant $g_{\Lambda N \kappa}=-8.312$ is taken from Nijmegen model NSC97a \cite{Stoks:1999}, determined by a fit to the $\Lambda N-\Sigma N$ scattering data.

The effective Lagrangians for hadronic vertices including nucleon resonances are
\begin{eqnarray}
{\cal L}_{R\Lambda K^*}^{1/2\pm} &=& -\, \frac{g_{R\Lambda K^*}}{2M_N}\bar{R}\Gamma^{(\mp)} \left\{ \left[ \left(
\frac{\gamma_\mu\partial^2}{M_R\mp M_N} \pm i\partial_\mu \right)  \right.\right. \nonumber \\
&&  -\left.\left. \frac{f_{R\Lambda K^*}}{g_{R\Lambda K^*}}\sigma_{\mu\nu}\partial^\nu \right] {K^*}^\mu
\right\} \Lambda   + \hc, \\[6pt]
{\cal L}_{R\Lambda {K^*}}^{3/2\pm} &=& -\, i\frac{g_{R\Lambda {K^*}}^{(1)}}{2M_N}\bar{R}_\mu \gamma_\nu \Gamma^{(\pm)}{K^*}^{\mu \nu} \Lambda  \nonumber \\
&& +\, \frac{g_{R\Lambda {K^*}}^{(2)}}{\left(2M_N\right)^2}\bar{R}_\mu \Gamma^{(\pm)}{K^*}^{\mu \nu}\partial_\nu \Lambda  \nonumber \\
&& \mp\, \frac{g_{R\Lambda {K^*}}^{(3)}}{\left(2M_N\right)^2}\bar{R}_\mu \Gamma^{(\pm)}\left(\partial_\nu {K^*}^{\mu \nu}\right) \Lambda   + \hc, \\[6pt]
{\cal L}_{R\Lambda {K^*}}^{5/2\pm} &= & \frac{g_{R\Lambda {K^*}}^{(1)}}{\left(2M_N\right)^2}\bar{R}_{\mu \alpha}\gamma_\nu \Gamma^{(\mp)}\left(\partial^{\alpha} K^{* \mu \nu}\right) \Lambda   \nonumber \\
&& \pm\,  i\frac{g_{R\Lambda K^*}^{(2)}}{\left(2M_N\right)^3}\bar{R}_{\mu \alpha} \Gamma^{(\mp)}\left(\partial^\alpha {K^*}^{\mu \nu}\right)\partial_\nu \Lambda  \nonumber \\
&& \mp\,  i\frac{g_{R\Lambda {K^*}}^{(3)}}{\left(2M_N\right)^3}\bar{R}_{\mu \alpha} \Gamma^{(\mp)} \left(\partial^\alpha \partial_\nu {K^*}^{\mu \nu}\right) \Lambda  \nonumber \\
&& +\, \hc, \\[6pt]
 {\cal L}_{R\Lambda {K^*}}^{7/2\pm} &=&  i\frac{g_{R\Lambda {K^*}}^{(1)}}{\left(2M_N\right)^3}\bar{R}_{\mu \alpha \beta}\gamma_\nu \Gamma^{(\pm)} \left(\partial^{\alpha}\partial^{\beta} {K^*}^{\mu \nu}\right) \Lambda   \nonumber  \\
&& -\, \frac{g_{R\Lambda {K^*}}^{(2)}}{\left(2M_N\right)^4}\bar{R}_{\mu \alpha \beta} \Gamma^{(\pm)} \left(\partial^\alpha \partial^\beta {K^*}^{\mu \nu}\right) \partial_\nu \Lambda   \nonumber  \\
 && \pm\, \frac{g_{R\Lambda {K^*}}^{(3)}}{\left(2M_N\right)^4}\bar{R}_{\mu \alpha \beta} \Gamma^{(\pm)} \left(\partial^\alpha \partial^\beta \partial_\nu   {K^*}^{\mu \nu}\right) \Lambda  \nonumber \\
&& +\, \hc.
\end{eqnarray}
In the present work, the coupling constant $f_{R\Lambda K^*}$ in ${\cal L}_{R\Lambda K^*}^{1/2\pm}$ is set to be zero, and the $g_{R\Lambda {K^*}}^{(2)}$ and $g_{R\Lambda {K^*}}^{(3)}$ terms in ${\cal L}_{R\Lambda {K^*}}^{3/2\pm}$, ${\cal L}_{R\Lambda {K^*}}^{5/2\pm}$ and ${\cal L}_{R\Lambda {K^*}}^{7/2\pm}$ are ignored for the sake of simplicity. These terms have been checked and found to be insensitive to the reaction amplitudes in the present investigation. The parameters $g_{R\Lambda K^*}$ and $g_{R\Lambda {K^*}}^{(1)}$ are fit parameters. Actually, only the products of the electromagnetic couplings and the hadronic couplings of nucleon resonances are relevant to the reaction amplitudes, and these products are what we really fit in practice.

The effective Lagrangian for the Kroll-Ruderman term of $\gamma N\to \Lambda K^*$ reads
\begin{eqnarray}
{\cal L}_{\gamma N \Lambda {K^*}} &=&  -\, i  g_{\Lambda N {K^*}} \frac{\kappa_{\Lambda N {K^*}}}{2M_N} \bar{\Lambda} \sigma^{\mu \nu} A_\nu  \hat{Q}_{K^*} K^*_\mu N \nonumber \\
&& +\, \hc,   \label{eq:L_gLNKst}
\end{eqnarray}
with $\hat{Q}_{K^*}$ being the electric charge operator of the outgoing $K^*$ meson. This interaction Lagrangian is obtained by the minimal gauge substitution $\partial_\mu \to {\cal D}_\mu\equiv \partial_\mu-i\hat{Q}_{K^*}A_\mu$ in the $\Lambda N K^*$ interaction Lagrangian of Eq.~(\ref{eq:L_LNKst}). The couplings $g_{\Lambda N {K^*}}$ and $\kappa_{\Lambda N {K^*}}$ have been given in Eqs.~(\ref{g_LNKst}) and (\ref{kappa_LNKst}).

\subsection{Resonance propagators}

In principle, an energy-dependent width of resonance is more realistic than a constant value multiplied by a step function. However, as discussed in Ref.~\cite{Nakayama:2006etap}, the cross section data alone are usually insensitive to the energy dependence of the resonance width. For the reaction of $\gamma N \to K^* \Lambda$, so far we only have the differential cross section data while the data for spin observables are not available. Hence, it is justified to treat the resonance width as a constant instead of a complex energy-dependent function for the sake of simplicity.

For spin-$1/2$ resonance propagator, we use the ansatz
\begin{equation}
S_{1/2}(p) = \frac{i}{\slashed{p} - M_R + i \Gamma/2},
\end{equation}
where $M_R$ and $\Gamma$ are the mass and width of resonance $R$ with four-momentum $p$, respectively.

Following Refs.~\cite{Behrends:1957,Fronsdal:1958,Zhu:1999}, the prescriptions of the propagators for resonances with spin-$3/2$, -$5/2$ and -$7/2$ are
\begin{eqnarray}
S_{3/2}(p) &=&  \frac{i}{\slashed{p} - M_R + i \Gamma/2} \left( \tilde{g}_{\mu \nu} + \frac{1}{3} \tilde{\gamma}_\mu \tilde{\gamma}_\nu \right),  \\[6pt]
S_{5/2}(p) &=&  \frac{i}{\slashed{p} - M_R + i \Gamma/2} \,\bigg[ \, \frac{1}{2} \big(\tilde{g}_{\mu \alpha} \tilde{g}_{\nu \beta} + \tilde{g}_{\mu \beta} \tilde{g}_{\nu \alpha} \big)  \nonumber \\
&& -\, \frac{1}{5}\tilde{g}_{\mu \nu}\tilde{g}_{\alpha \beta}  + \frac{1}{10} \big(\tilde{g}_{\mu \alpha}\tilde{\gamma}_{\nu} \tilde{\gamma}_{\beta} + \tilde{g}_{\mu \beta}\tilde{\gamma}_{\nu} \tilde{\gamma}_{\alpha}  \nonumber \\
&& +\, \tilde{g}_{\nu \alpha}\tilde{\gamma}_{\mu} \tilde{\gamma}_{\beta} +\tilde{g}_{\nu \beta}\tilde{\gamma}_{\mu} \tilde{\gamma}_{\alpha} \big) \bigg], \\[6pt]
S_{7/2}(p) &=&  \frac{i}{\slashed{p} - M_R + i \Gamma/2} \, \frac{1}{36}\sum_{P_{\mu} P_{\nu}} \bigg( \tilde{g}_{\mu_1 \nu_1}\tilde{g}_{\mu_2 \nu_2}\tilde{g}_{\mu_3 \nu_3} \nonumber \\
&& -\, \frac{3}{7}\tilde{g}_{\mu_1 \mu_2}\tilde{g}_{\nu_1 \nu_2}\tilde{g}_{\mu_3 \nu_3} + \frac{3}{7}\tilde{\gamma}_{\mu_1} \tilde{\gamma}_{\nu_1} \tilde{g}_{\mu_2 \nu_2}\tilde{g}_{\mu_3 \nu_3} \nonumber \\
&& -\, \frac{3}{35}\tilde{\gamma}_{\mu_1} \tilde{\gamma}_{\nu_1} \tilde{g}_{\mu_2 \mu_3}\tilde{g}_{\nu_2 \nu_3} \bigg),  \label{propagator-7hf}
\end{eqnarray}
where
\begin{eqnarray}
\tilde{g}_{\mu \nu} &=& -\, g_{\mu \nu} + \frac{p_{\mu} p_{\nu}}{M_R^2}, \\[6pt]
\tilde{\gamma}_{\mu} &=& \gamma^{\nu} \tilde{g}_{\nu \mu} = -\gamma_{\mu} + \frac{p_{\mu}\slashed{p}}{M_R^2},
\end{eqnarray}
and the summation over $P_\mu$ $\left(P_\nu\right)$ in Eq.~(\ref{propagator-7hf}) goes over the $3!=6$ possible permutations of the indices $\mu_1\mu_2\mu_3$ $\left(\nu_1\nu_2\nu_3\right)$. These high-spin resonance propagators and their variations have been applied with success in a number of  resonance studies \cite{Man:2011,Kim:2011,Huang:2013,Jackson:2015,Kim:2014}.

\subsection{Form factors}

\begin{table*}[tb]
\caption{\label{Table:para} Model parameters in five different fits. Here $\beta_{\Lambda K^*}$ is the branching ratio for resonance decay to $\Lambda K^*$, and $A_{1/2}$, $A_{3/2}$ are helicity amplitudes for resonance radiative decay to $\gamma p$. 
For the definition of other parameters, see Sec.~\ref{Sec:formalism}. The stars below resonance names denote the overall status of these resonances evaluated by the most recent PDG \cite{Patrignani:2016}. The numbers in brackets below the resonance masses and widths represent the corresponding values estimated by the most recent PDG \cite{Patrignani:2016}.}
\renewcommand{\arraystretch}{1.4}
\smallskip
\begin{tabular*}{\textwidth}{@{\extracolsep\fill}lrrrrr}
\hline\hline
Model          &        I      &      II       &       III     &       IV     &      V     \\
$\chi^2/N$   &  $1.35$  &  $1.79$  &  $1.85$  &  $2.09$  &  $2.18$  \\
\hline
$g^{(1)}_{\Sigma^* \Lambda \gamma}$  & $0.74\pm 0.16$ & $-0.90\pm 0.17$  & $-0.87\pm 0.14$  &  $-0.60\pm 0.18$  &  $-0.22\pm 0.16$  \\
$\Lambda_K$ [MeV]   & $1000\pm 6$    &   $1019\pm 4$   & $993\pm 7$   &   $1030\pm 3$   &  $1018\pm 4$    \\
\hline
$N^*$ Name  &  $N(2060){5/2}^-$  &  $N(2060){5/2}^-$  &  $N(2060){5/2}^-$  &  $N(2060){5/2}^-$  &  $N(2060){5/2}^-$    \\[-3pt]
                      &     $**$    &     $**$    &      $**$    &      $**$    &      $**$       \\
$M_R$ [MeV]              & $2033\pm 2$     & $2009\pm 5$     & $2032\pm 3$  & $2043\pm 4$  & $2038\pm 3$  \\
$\Gamma_R$ [MeV]   & $65\pm 4$         &  $213\pm 20$        &    $81\pm 8$   &     $202\pm 16$   & $77\pm 8$   \\
$\Lambda_R$ [MeV]   & $1188\pm 20$   &  $965\pm 16$  & $1126\pm 12$ & $889\pm 13$ & $981\pm 22$  \\
$\sqrt{\beta_{\Lambda K^*}}A_{1/2}$ [$10^{-3}$\,GeV$^{-1/2}$]  & $0.69\pm 0.06$  & $0.03\pm 0.01$  &  $0.33\pm 0.03$  &  $0.60\pm 0.06$ & $-0.21\pm 0.02$  \\
$\sqrt{\beta_{\Lambda K^*}}A_{3/2}$ [$10^{-3}$\,GeV$^{-1/2}$]  & $-1.39\pm 0.13$  &  $-0.10\pm 0.01$  &   $-1.10\pm 0.10$  & $-1.94\pm 0.19$ & $-1.56\pm 0.15$   \\
\hline
$N^*$ Name  &  $N(2000){5/2}^+$  &  $N(2040){3/2}^+$  &  $N(2120){3/2}^-$  &  $N(2190){7/2}^-$  &  $N(2100){1/2}^+$   \\[-3pt]
                      &     $**$    &     $*$    &      $**$    &      $*\!*\!**$    &      $*$       \\
$M_R$ [MeV]              & $2115\pm 22$  & $2200\pm 62$ & $2203\pm 9$  & $2243\pm 6$  & $2100\pm 15$  \\
                                    &                          &                         & $[\approx 2120]$  &  $[2100 \sim 2200]$  &  $[\approx 2100]$   \\
$\Gamma_R$ [MeV]   & $450\pm 10$    &  $540\pm 7$   &  $433\pm 33$   &     $450\pm 33$   & $450\pm 9$   \\
                                    &                          &                         &                        &  $[300 \sim 700]$  &                          \\
$\Lambda_R$ [MeV]   & $1644\pm 21$   &  $1564\pm 36$  & $1726\pm 58$ & $936\pm 13$ & $1431\pm 31$  \\
$\sqrt{\beta_{\Lambda K^*}}A_{1/2}$ [$10^{-3}$\,GeV$^{-1/2}$]  & $-2.87\pm 0.81$  & $3.12\pm 0.85$  &  $4.53\pm 0.38$  &  $5.21\pm 0.33$ & $-7.22\pm 1.40$  \\
$\sqrt{\beta_{\Lambda K^*}}A_{3/2}$ [$10^{-3}$\,GeV$^{-1/2}$]  & $-1.04\pm 0.29$  &  $7.87\pm 2.13$  &   $7.84\pm 0.65$  & $3.71\pm 0.24$ &    \\
\hline\hline
\end{tabular*}
\end{table*}

Each hadronic vertex obtained from the Lagrangians given in Sec.~\ref{Sec:Lagrangians} is accompanied with a phenomenological form factor to parametrize the structure of the hadrons and to normalize the behavior of the production amplitude. Following Refs.~\cite{Kim:2011,Kim:2014}, for intermediate baryon exchange we take the form factor as
\begin{eqnarray}
f_B(p^2) = \left(\frac{\Lambda_B^4}{\Lambda_B^4+\left(p^2-M_B^2\right)^2}\right)^n,  \label{eq:ff_B}
\end{eqnarray}
where $p$ denotes the four-momentum of the intermediate baryon, the exponent $n$ is taken to be $2$ for all baryon exchanges, and the cutoff $\Lambda_B$ is taken to be $900$ MeV for all $N$, $\Lambda$, $\Sigma$ and $\Sigma^*$ exchanges \cite{Kim:2011,Kim:2014}. For the $s$-channel resonance exchanges, the cutoffs are treated as fit parameters. For intermediate meson exchange, we take the form factor as
\begin{eqnarray}
f_M(q^2) = \left(\frac{\Lambda_M^2-M_M^2}{\Lambda_M^2-q^2}\right)^m, \label{eq:ff_M}
\end{eqnarray}
where $q$ represents the four-momentum of the intermediate meson, the exponent $m$ is taken to be $2$  for all meson exchanges, $M_M$ and $\Lambda_M$ designate the mass and cutoff mass of exchanged meson $M$. We choose $M_\kappa=800$ MeV, and for other exchanged mesons, the experimental values are used for their masses. The cutoffs $\Lambda_{K^*}=900$ MeV and $\Lambda_\kappa=1100$ MeV are adopted in the present work which are also taken from Ref.~\cite{Kim:2014}. The cutoff $\Lambda_K$ is treated as a free parameter and will be determined by a fit to the experimental differential cross section data.

Note that the gauge-invariance feature of our photoproduction amplitude is independent of the specific form of the form factors.

\section{Results and discussion}   \label{Sec:results}

As mentioned in Sec.~\ref{Sec:intro}, the work of Ref.~\cite{Kim:2014} presents so far the only detailed theoretical analysis of the most recent high-statistics differential cross section data from CLAS \cite{Tang:2013} for the $K^{*+}\Lambda$ photoproduction reaction. There, four nucleon resonances, namely $N(2000){5/2}^+$, $N(2060){5/2}^-$, $N(2120){3/2}^-$ and $N(2190){7/2}^-$, have been considered with the parameters of $N(2000){5/2}^+$ and $N(2060){5/2}^-$ taken from Ref.~\cite{Anisovich:2013}, the parameters of $N(2190){7/2}^-$ taken from a relativistic quark model calculation \cite{Capstick:1992}, and the parameters of $N(2120){3/2}^-$ determined by a fit to the experimental data. It was found that the $N(2120){3/2}^-$ and $N(2190){7/2}^-$ resonances are essential in describing the measured cross section data. The $N(2060){5/2}^-$ resonance was found to have a relative small but still noticeable contribution, while the $N(2000){5/2}^+$ was found to be negligible in this reaction.

\begin{figure*}[tbp]
\includegraphics[width=0.65\textwidth]{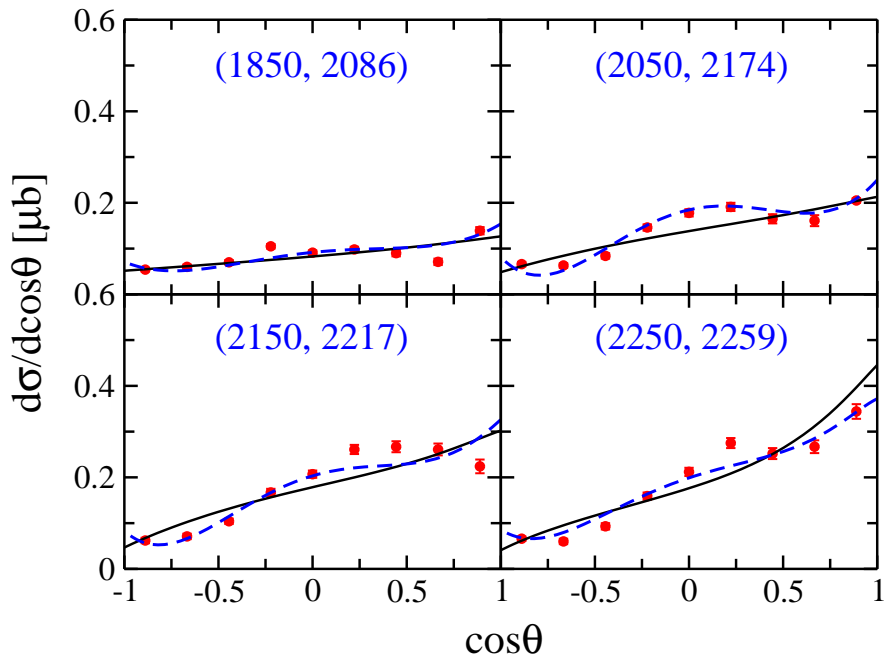}
\caption{(Color online) Differential cross sections for $\gamma p\to K^{*+}\Lambda$ as a function of $\cos\theta$ in the center-of-mass frame in the near threshold region. The black solid lines correspond to the fit result including the $N(2000)5/2^+$ and $N(2040)3/2^+$ resonances with $\chi^2/N=2.50$. The blue dashed lines represent the results from model V with $\chi^2/N=2.18$. The scattered symbols are the most recent data from CLAS Collaboration \cite{Tang:2013}. The numbers in parentheses denote the photon laboratory incident energy (left number) and the total center-of-mass energy of the system (right number), in MeV.}
\label{fig:other_Rs}
\end{figure*}

In the present work, we adopt a rather different strategy than Ref.~\cite{Kim:2014} for investigating the roles of nucleon resonances in the $\gamma p \to K^{*+} \Lambda$ reaction. That is, in addition to the Born term which is composed of the $t$-channel $K$, $\kappa$, $K^*$ exchanges, the $u$-channel $\Lambda$, $\Sigma$, $\Sigma^*(1385)$ exchanges, the $s$-channel $N$ exchange, and the generalized contact current as illustrated in Fig.~\ref{FIG:feymans}, we introduce the $s$-channel nucleon resonances as few as possible in constructing the reaction amplitudes in order to achieve a satisfactory fit to the high-statistics differential cross section data from CLAS \cite{Tang:2013}. In practice, we allow in our model all those six resonances near the $K^* \Lambda$ threshold, namely, $N(2000){5/2}^+$, $N(2040){3/2}^+$, $N(2060){5/2}^-$, $N(2100){1/2}^+$, $N(2120){3/2}^-$, and $N(2190){7/2}^-$. After numerous trials with the inclusion of different number of nucleon resonances and different combinations, we found that, if only one resonance is included, the $\chi^2$ per data point, $\chi^2/N$, are all larger than $3$. The quality of the corresponding fit results are then found to be significantly poor, and thus, they are treated as unacceptable fit results. If two resonances are included, it is found that there are five possible sets of resonance combinations which result in fits with $\chi^2/N \lesssim 2.18$ and these fits are visually in good agreement with the data. All these five sets require a common resonance, $N(2060)5/2^-$. The other resonance is one of the $N(2000){5/2}^+$, $N(2040){3/2}^+$, $N(2100){1/2}^+$, $N(2120){3/2}^-$, and $N(2190){7/2}^-$ resonances. The other combinations of two resonances all ended up in $\chi^2/N \gtrsim 2.50$, noticeably of inferior quality even with the naked eye. Hence, they are not considered as acceptable fit results. A comparison of the fit results with two resonances, one with $\chi^2/N = 2.18$ (accepted result corresponding to model V as it will be discussed later) and another with $\chi^2/N=2.50$ (unaccepted result with the resonances $N(2000)5/2^+$ and $N(2040)3/2^+$), is shown in Fig.~\ref{fig:other_Rs}. The difference in the fit quality is clearly seen even with the naked eye. Now, if three resonances are considered, the $\chi^2/N$ improves only slightly compared with that with two resonances. We thus conclude that one needs at least two resonances to obtain a reasonable fit of the cross section data for $\gamma p\to K^{*+} \Lambda$ in the present approach.

\begin{figure*}[tbp]
\includegraphics[width=0.88\textwidth]{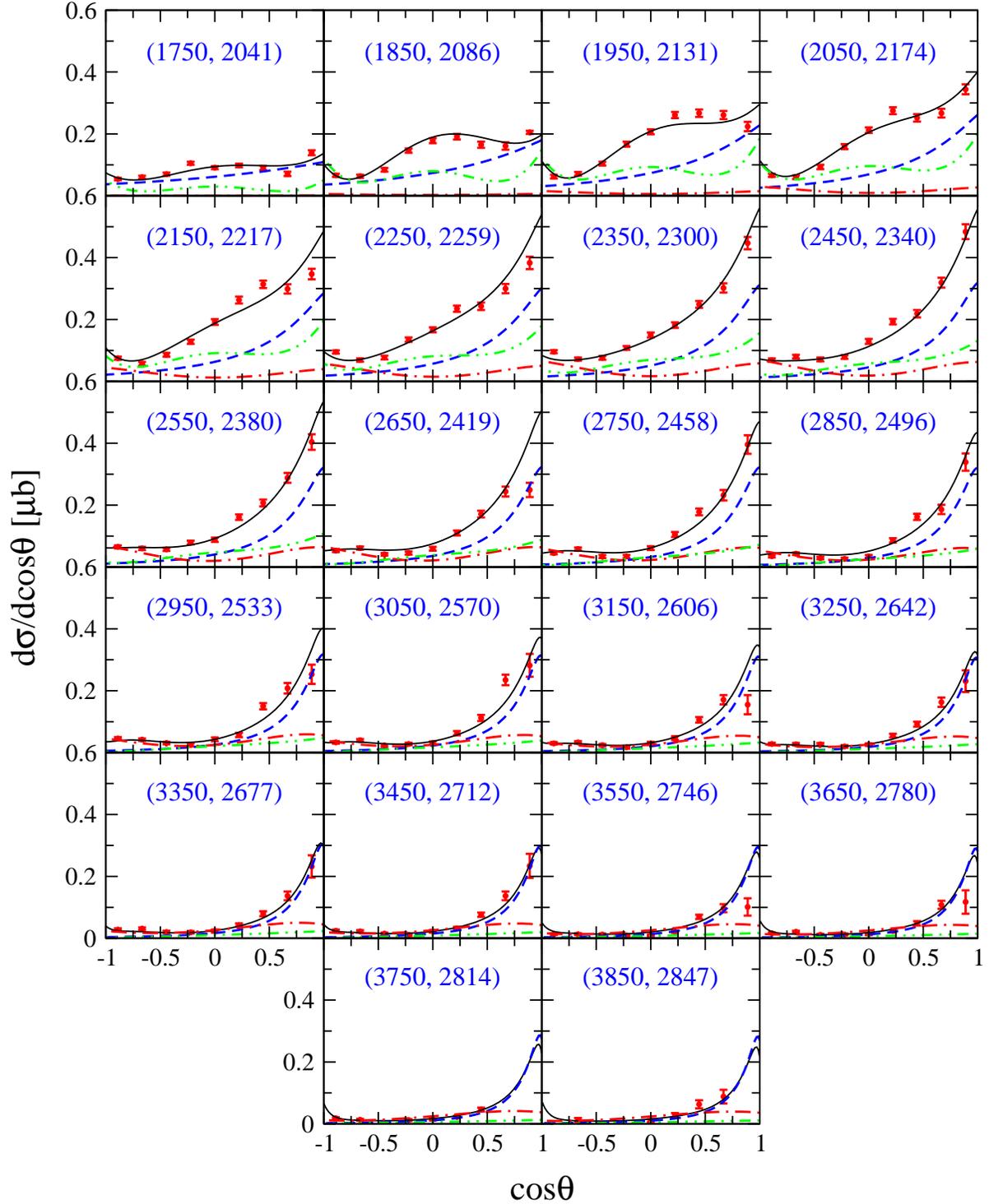}
\caption{(Color online) Differential cross sections for $\gamma p \to K^{*+}\Lambda$ as a function of $\cos\theta$ from model I (black solid lines). The scattered symbols denote the CLAS data \cite{Tang:2013}. The blue dashed, green dash-double-dotted, and magenta dash-dotted lines represent the individual contributions from $K$, $N(2060){5/2}^-$ and $N(2000){5/2}^+$ exchanges, respectively. The photon incident energy binning is $100$ MeV. The numbers in parentheses denote the centroid value of the photon laboratory incident energy (left number) and the corresponding total center-of-mass energy of the system (right number), in MeV.}
\label{fig:fit23}
\end{figure*}

\begin{figure*}[tbp]
\includegraphics[width=0.88\textwidth]{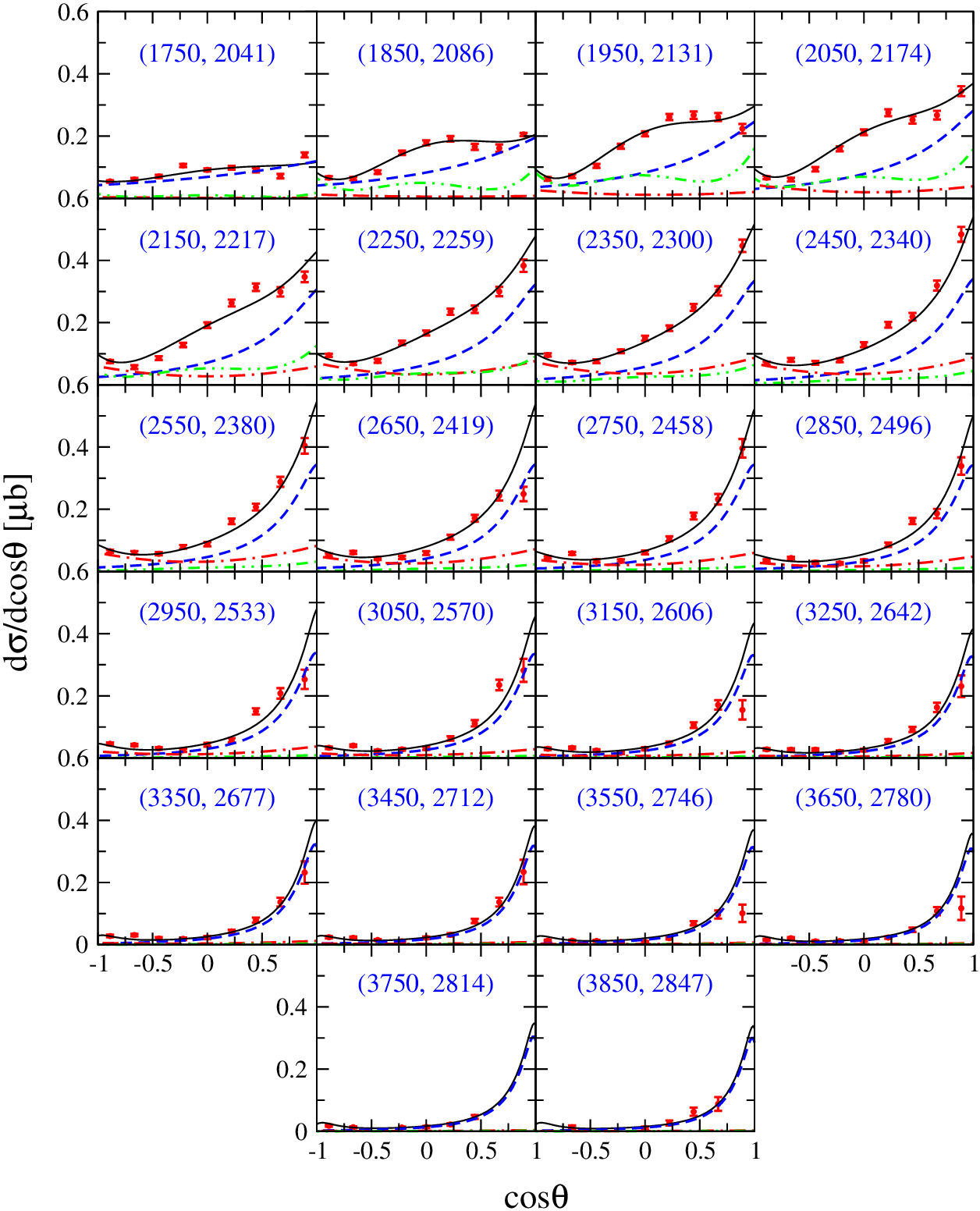}
\caption{(Color online) Differential cross sections for $\gamma p \to K^{*+}\Lambda$ as a function of $\cos\theta$ from model II (black solid lines). The notations are the same as in Fig.~\ref{fig:fit23} except that now the magenta dash-dotted lines represent the contribution from $s$-channel $N(2040){3/2}^+$ exchange.}
\label{fig:fit35}
\end{figure*}

\begin{figure*}[tbp]
\includegraphics[width=0.88\textwidth]{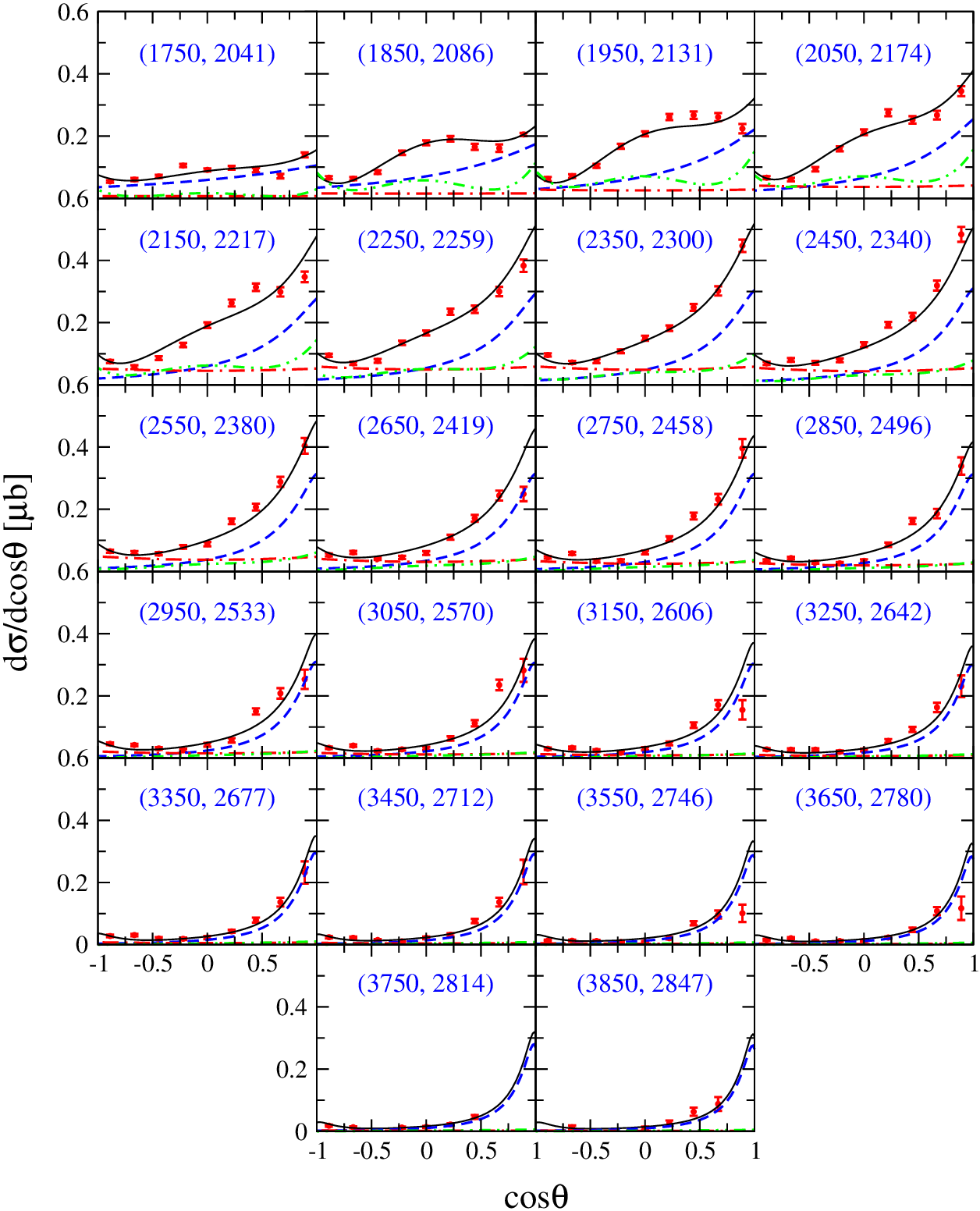}
\caption{(Color online) Differential cross sections for $\gamma p \to K^{*+}\Lambda$ as a function of $\cos\theta$ from model III (black solid lines). The notations are the same as in Fig.~\ref{fig:fit23} except that now the magenta dash-dotted lines represent the contribution from $s$-channel $N(2120){3/2}^-$ exchange.}
\label{fig:fit13}
\end{figure*}

\begin{figure*}[tbp]
\includegraphics[width=0.88\textwidth]{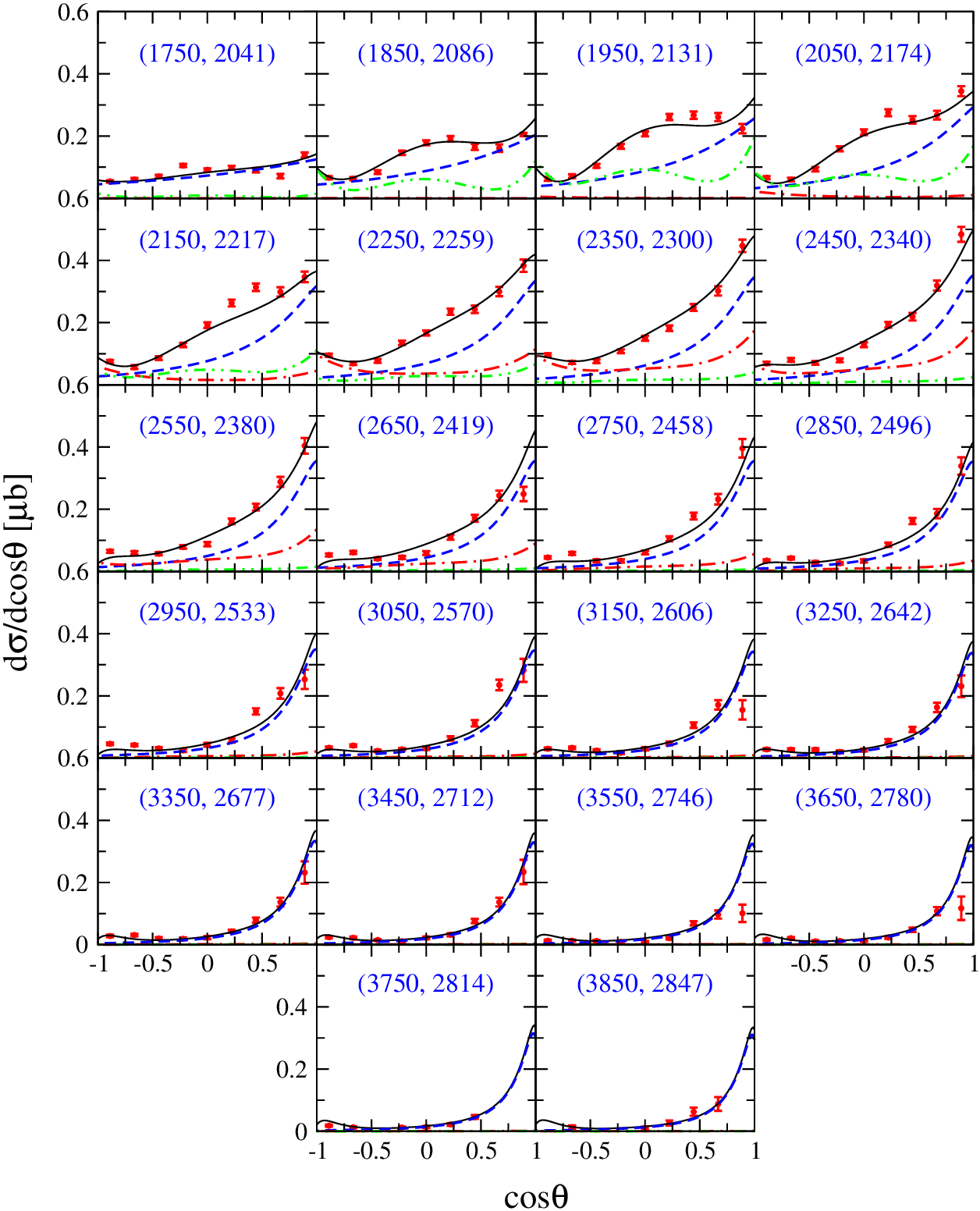}
\caption{(Color online) Differential cross sections for $\gamma p \to K^{*+}\Lambda$ as a function of $\cos\theta$ from model IV (black solid lines). The notations are the same as in Fig.~\ref{fig:fit23} except that now the magenta dash-dotted lines represent the contribution from $s$-channel $N(2190){7/2}^-$ exchange.}
\label{fig:fit34}
\end{figure*}

\begin{figure*}[tbp]
\includegraphics[width=0.88\textwidth]{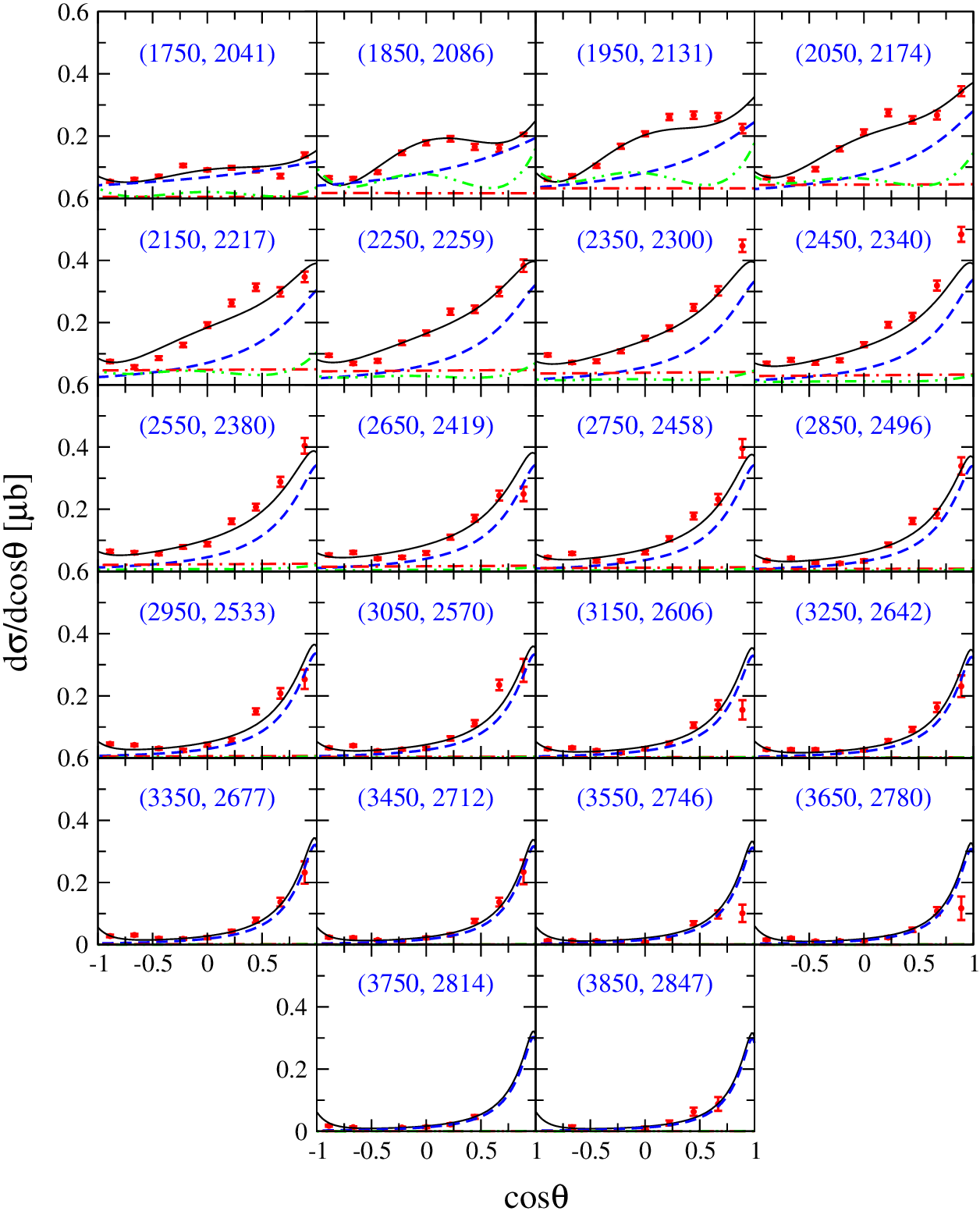}
\caption{(Color online) Differential cross sections for $\gamma p \to K^{*+}\Lambda$ as a function of $\cos\theta$ from model V (black solid lines). The notations are the same as in Fig.~\ref{fig:fit23} except that now the magenta dash-dotted lines represent the contribution from $s$-channel $N(2100){1/2}^+$ exchange.}
\label{fig:fit36}
\end{figure*}

\begin{figure}[tbp]
\includegraphics[width=0.81\columnwidth]{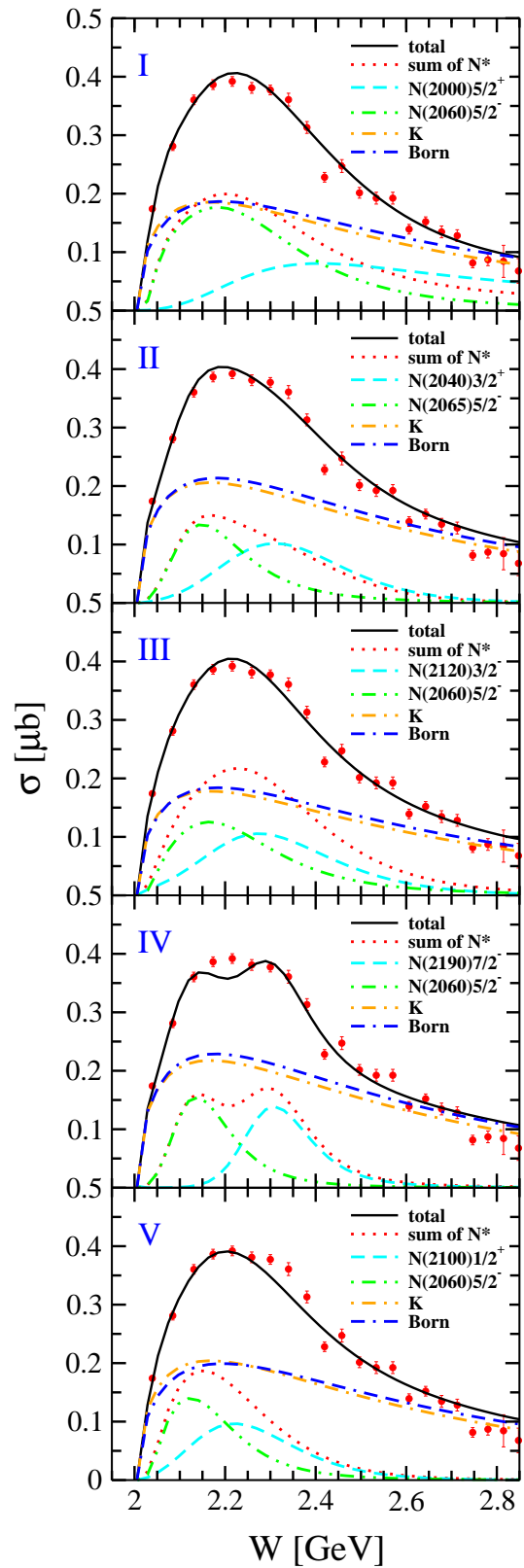}
\caption{(Color online) Total cross sections with individual (resonance, Born term, $K$) contributions for $\gamma p \to K^{*+}\Lambda$. The panels from top to bottom correspond to the results of mode I-V, as indicated. The data are from CLAS \cite{Tang:2013} but not included in the fit.  }
\label{fig:total_cro_sec}
\end{figure}

We now turn to the discussion of the details of our analysis of the data with two nucleon resonances included. As mentioned above, in this case there are five different sets of the resonance combination which result in fits describing the differential cross section data of $K^{*+}$ photoproduction reaction satisfactorily according to our criterium of $\chi^2/N < 2.5$. The fitted values of all the adjustable parameters in those five models are listed in Table~\ref{Table:para}. There, the stars below resonance names denote the overall status of these resonances evaluated by the most recent review by the PDG \cite{Patrignani:2016}, and the numbers in brackets below the resonance masses and widths represent the corresponding estimates given by the PDG. The uncertainties in the resulting parameters are estimates arising from the uncertainties (error bars) associated with the fitted experimental differential cross section data points. For each resonance, apart from its mass, total width and cutoff parameter in the form factor, the table also shows the corresponding reduced helicity amplitudes $\sqrt{\beta_{\Lambda K^*}}A_j$, where $\beta_{\Lambda K^*}$ denotes the branching ratio to the decay channel $\Lambda K^*$ and $A_j$ stands for the helicity amplitude with spin $j$. We mention that only the product of these two quantities can be well constrained in the present work as the $s$-channel (resonance) amplitudes are sensitive only to the product of the hadronic and electromagnetic coupling constants, a feature common to single channel calculations.  Following Ref.~\cite{Huang:2013}, here we have assumed a radiative branching ratio of $\beta_{p\gamma}=0.2\%$ for all the resonances to calculate the corresponding helicity amplitudes from the associated product of the hadronic and electromagnetic coupling constants. It is seen from Table~\ref{Table:para} that the coupling $g^{(1)}_{\Sigma^* \Lambda \gamma}$ varies much from one model to another. This is simply because the $u$-channel $\Sigma^*(1385)$ exchange has negligible contribution to the reaction $\gamma p\to K^{*+} \Lambda$ (cf. Figs.~\ref{fig:fit23}-\ref{fig:fit36} and Fig.~\ref{fig:total_cro_sec} discussed later in this section).  The fitted values of $\Lambda_K$, the cutoff parameter in the $K$-meson exchange contribution, are very close to each other in models I-V --- they are all around $1.0$ GeV. This value is determined mainly by the data in the high energy region, where the $K$ meson exchange dominates the whole amplitude of this reaction (cf. Figs.~\ref{fig:fit23}-\ref{fig:fit36} and Fig.~\ref{fig:total_cro_sec}). The fitted values of the mass of $N(2060)5/2^-$ from various models are also very close to each other, while those of its decay width are not.  In each of the models I-V, the value of the fitted mass of the other resonance is compatible with that quoted in PDG \cite{Patrignani:2016}. The fit result for the width of the four-star $N(2190)7/2^-$ is compatible with the PDG estimate which has a large range. The widths of the $N(2000)5/2^+$,  $N(2040)3/2^+$, $N(2120)3/2^-$, and $N(2100)1/2^+$ resonances obtained in the present work are somewhat larger than those obtained in the other analyses listed in the PDG. We note that the reduced helicity amplitudes for $N(2060){5/2}^-$ corresponding to model II are much smaller than those corresponding to the other models. This is caused by the smaller value of the resulting resonance mass of $2009 \pm 5$ MeV for model II, leading to a much smaller branching ratio $\beta_{\Lambda K^*}$. Note that the $R \to \Lambda K^*$ decay threshold is $ 2007$ MeV, thus, for model II, the $N(2060)5/2^-$ resonance is only 2 MeV above the threshold. We will discuss further details below in connection with the differential cross section results shown in Figs.~\ref{fig:fit23}-\ref{fig:fit36} and the total cross section results shown in Fig.~\ref{fig:total_cro_sec}.

The results for differential cross sections corresponding to the model
parameters (I-V) listed in Table~\ref{Table:para} are shown in
Figs.~\ref{fig:fit23}-\ref{fig:fit36}, where the contributions from the
$t$-channel $K$-meson exchange and from the individual $s$-channel resonance
exchanges are also presented. In these figures, the black solid lines
correspond to the total contribution (coherent sum of all the individual
contributions), the blue dashed lines represent the contribution from the
$t$-channel $K$-meson exchange, the green dash-double-dotted lines, the
$s$-channel $N(2060)5/2^-$ exchange, and the magenta dash-dotted lines denote
the contribution from the other $s$-channel resonance exchange in the
corresponding models. The contributions from the other terms are too small to
be clearly seen with the scale used, and thus, they are not plotted. One sees
from Figs.~\ref{fig:fit23}-\ref{fig:fit36} that the overall description of the
CLAS high-statistics angular distribution data is fairly satisfactory in all of
the five models. In particular, the angular dependence of the differential
cross sections near the $\Lambda K^*$ threshold is qualitatively in good
agreement with the data, much better than the description of
Ref.~\cite{Kim:2014} (cf. Fig.~\ref{Fig:status}). This can be understood if one
notices that in all our models I-V, there is a significant contribution from
the $N(2060){5/2}^-$ resonance in the low energy region that is responsible for
reproducing the observed shape of the angular distribution through an
interference with the large $K$-meson exchange contribution near the $\Lambda
K^*$ threshold. In contrast, Ref.~\cite{Kim:2014} has a rather small
contribution of $N(2060){5/2}^-$ and clear discrepancies are seen in its
description of the near threshold differential cross section data (cf.
Fig.~\ref{Fig:status}). We note that the contributions other than the
$N(2060){5/2}^-$ resonance and $K$-meson exchange are practically negligible in
these low-energy region. In particular, when the $N(2000){5/2}^+$ contribution
is switched off, one obtains the (full contribution) results that practically
coincide with the $K$-meson exchange contribution alone. The above
considerations explain, in particular, why the mass of $N(2060){5/2}^-$ is
fairly well constrained as can be seen from the resulting values in
Table~\ref{Table:para} for models I-V. The contribution of the $N(2060){5/2}^-$
resonance in the high energy region is practically negligible. The other
resonance contribution in each of the models I-V is considerable and mainly
around $W \sim 2.2$ GeV, and it becomes very small as the energy goes down
approaching the $\Lambda K^*$ threshold or goes up to higher energies. The role
of the individual resonances as a function of energy can be better seen in the
total cross section section (cf. Fig.~\ref{fig:total_cro_sec}). The
contribution of the $K$ meson exchange is seen to be very important in the
whole energy region considered. Especially, it plays a crucial role in
reproducing the observed forward-peaked angular distribution at higher
energies. This is a general feature observed in many reactions at high
energies, where the $t$-channel mechanism accounts for the behavior of the
cross section at small $t$. This explains why the cutoff parameter values
($\Lambda_K$), which is the only adjustable parameter for $K$ meson exchange,
are close to each other in all of our models I-V. In other words, our $K$-meson
exchange contribution -- which practically exhausts the calculated non-resonant
background in the entire energy region -- is largely constrained by the data at
high energies. This leads to a much more unambiguous determination of the
resonance contributions in the present model. On the other hand, it is also
very interesting to see how the Regge trajectory description of the present
reaction would affect the strong angular dependence at very forward angles at
high energies exhibited by the $K$-meson exchange mechanism where no data exist
due to the limitations in the forward-angle acceptance of the CLAS detector
\cite{Tang:2013}. The gauge-invariant dynamical Regge approach put
forward in Ref.~\cite{Haberzettl:2015} seems well suited for this purpose;
however, this is left for a future investigation.

Before we leave the discussion of the differential cross section results, we mention that, although the present calculation describes the differential cross section data quite well overall and much better than any of the earlier calculations, the agreement with the data is not perfect. Indeed, the details of the observed angular behavior at $W=2.217$ GeV is not quite described by any of our models I-V. Also, our models show a slight tendency to miss the data at the neighboring energies of $W=2.174$ and $2.259$ GeV. As mentioned before, the inclusion of one more resonance didn't help improve much the fit quality.  A further investigation is required here.

\begin{figure}[tbp]
\includegraphics[width=0.81\columnwidth]{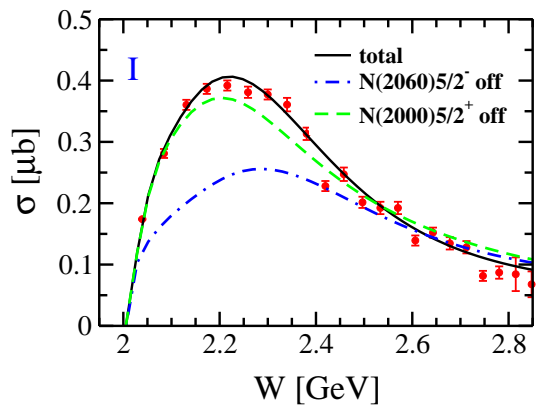}
\caption{(Color online) Same as in Fig.~\ref{fig:total_cro_sec} for model I. The blue dash-dotted line corresponds to the results with resonance $N(2060){5/2}^-$ switched off, while the green dashed line to those with $N(2000){5/2}^+$ switched off. The black solid line is the results of model I shown in Fig.~\ref{fig:total_cro_sec}. }
\label{fig:2on_off}
\end{figure}

\begin{figure}[htb]
\vglue 0.1cm
\includegraphics[width=0.81\columnwidth]{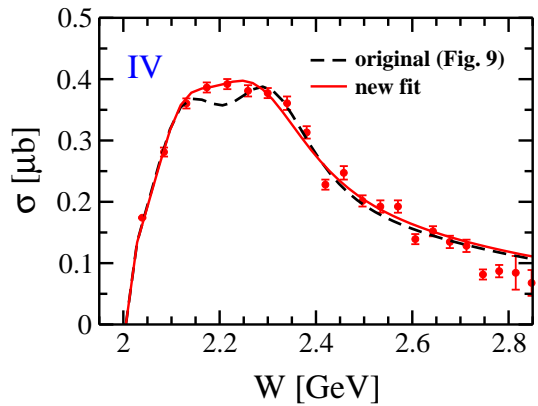}
\caption{(Color online) Same as in Fig.~\ref{fig:total_cro_sec} for model IV, except that it has been artificially forced to better reproduce the measured differential cross section at $W=2.217$ GeV (red solid line). The black dashed line is the results of model IV shown in Fig.~\ref{fig:total_cro_sec}.}
\label{fig:weight}
\end{figure}

\begin{figure*}[htb]
\includegraphics[width=0.67\textwidth]{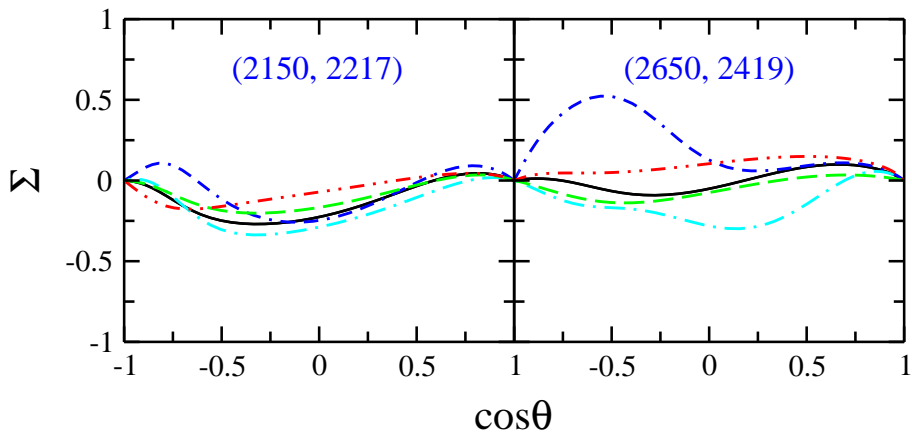}
\caption{(Color online) Photon beam asymmetries as functions of cosine of the $K^*$ emission angle $\theta$ in the center-of-mass system at two energies for the $\gamma p \to K^{*+} \Lambda$ reaction. The numbers in parentheses denote the photon laboratory incident energy (left number) and the total center-of-mass energy of the system (right number), in MeV. The blue double-dash-dotted curve, green dashed curve, black solid curve, cyan dash-dotted curve, and orange dash-double-dotted curve represent the predictions corresponding to the models I-V, respectively.}
\label{fig:beam_asy}
\end{figure*}

\begin{figure*}[htb]
\vglue 0.2cm
\includegraphics[width=0.67\textwidth]{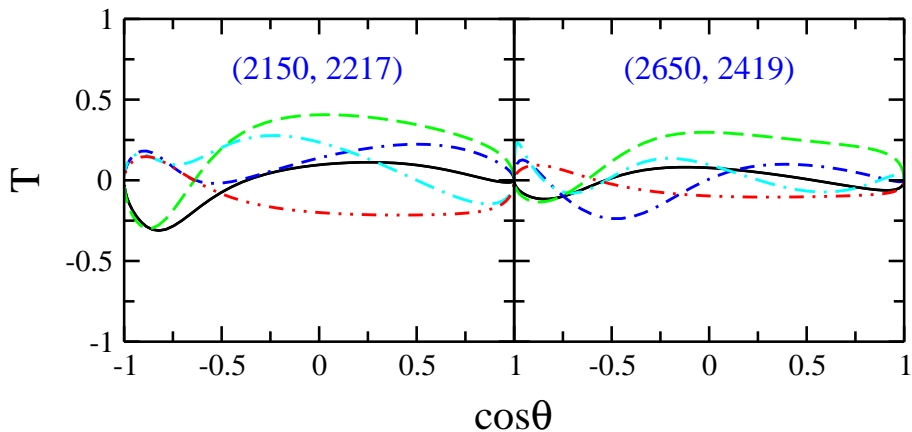}
\caption{(Color online) Same as in Fig.~\ref{fig:beam_asy} for target nucleon asymmetries.}
\label{fig:target_asy}
\end{figure*}

\begin{figure*}[htb]
\includegraphics[width=0.67\textwidth]{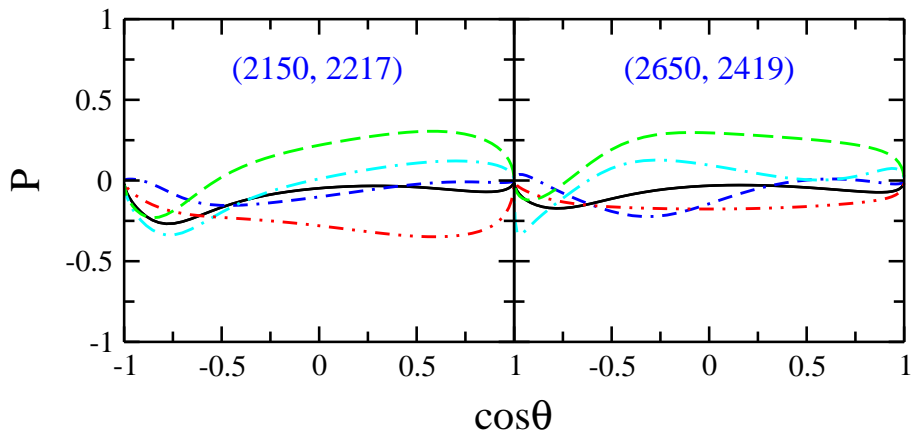}
\caption{(Color online) Same as in Fig.~\ref{fig:beam_asy} for recoil $\Lambda$ asymmetries.}
\label{fig:recoil_asy}
\end{figure*}

Figure~\ref{fig:total_cro_sec} shows the predicted total cross sections (black
solid line) together with the individual contributions from the $K$-exchange
(orange dash-dotted line), Born term (blue double-dash-dotted line) and the
resonances (green dash-double-dotted line and cyan dashed line) obtained by
integrating the corresponding results for differential cross sections from our
models I-V as shown in Figs.~\ref{fig:fit23}-\ref{fig:fit36}. We recall that
the Born term consists of the coherent sum of all the contributions other than
the $s$-channel resonance exchanges, i.e., the coherent sum of the $s$-channel
$N$ exchange, the $t$-channel $K$, $K^*$ and $\kappa$ exchanges, the
$u$-channel $\Lambda$, $\Sigma$ and $\Sigma^*(1385)$ exchanges, and the
generalized contact current. Note that the total cross section data are not
included in our fits. In this regard, it should be emphasized that the CLAS
total cross section data have been obtained by integrating the measured
differential cross sections and suffer from the limited forward-angle
acceptance of the CLAS detector \cite{Tang:2013}, especially at high energies
where the cross section exhibits a strong angular dependence at very forward
angles (cf. Figs.~\ref{fig:fit23}-\ref{fig:fit36}). The lack of differential
cross section data at very forward angles and high energies reflects in less
accurate CLAS total cross section data at these energies. Hence, caution must
be exercised when confronting these data with theoretical predictions. One sees
from Fig.~\ref{fig:total_cro_sec} that in all of our models I-V, the predicted
total cross sections are in fairly good agreement with the data over the entire
energy region considered, with one exception that, in model IV, the predicted
total cross sections exhibit a small valley structure around $W \sim 2.2$ GeV;
we come back to this point later. The $K$ meson exchange is seen to play an
important role in the whole energy region, especially at high energies. Its
contribution in all of the models I-V is more or less similar to each other, as
the only adjustable parameter for $K$ meson exchange, the cutoff mass
$\Lambda_K$ in the form factor, is constrained by the differential cross
section data at high energies, where this contribution practically dominates
this observable. We observe that our total cross section predictions exhibit a
slight tendency to overestimate the data at very high energies, although the
data are much less accurate and may suffer from the limitations in the CLAS
detector acceptance as mentioned above. Nevertheless, as mentions previously in
connection to the discussion of the differential cross sections, it would be
interesting to see how the Regge trajectory description of the present reaction
along the lines suggested in Ref.~\cite{Haberzettl:2015} would affect
the cross section at these high energies. The contributions from the
non-resonant terms other than the $K$-meson exchange are negligible, as can be
seen by comparing the Born term (red dash-double-dotted line) with the $K$
exchange (blue short dashed line) contribution. The negligible contribution of
the $u$-channel $\Sigma^*(1385)$ exchange -- which is a part of the
non-resonant term -- explains why the coupling
$g^{(1)}_{\Sigma^*\Lambda\gamma}$ varies so much from one model to another in
Table~\ref{Table:para} with roughly similar fit qualities.

The broad bump exhibited by the total cross section is caused by the coherent
sum of the considered two resonances and $K$-meson exchange contributions. In
all of the models I-V, the sharp rise of the cross section from the threshold
up to $W \sim 2.15$ GeV is caused by the $K$-meson exchange and
$N(2060){5/2}^-$ resonance. The other resonance contributes mostly at higher
energies. These are better seen in Fig.~\ref{fig:2on_off}, where the effects of
the $N(2060){5/2}^-$ and $N(2000){5/2}^+$ on the total cross section are shown
by switching off these resonances one at a time. One clearly sees that the
$K$-meson exchange is responsible for the sharp raise of the total cross
section right from the threshold followed by the build up due to
$N(2060){5/2}^-$ as the energy increases up to $W \sim 2.1$ GeV. In
Ref.~\cite{Kim:2014}, the differential cross section near the $\Lambda K^*$
threshold is not well described (cf. Fig.~\ref{Fig:status}), and consequently,
the total cross section in this energy region is, to some extent,
underestimated. There, the broad bump is mainly described by the sum of the
contributions of $N(2120){3/2}^-$, $N(2060)5/2^-$, and $N(2190){7/2}^-$, and
the sharp rise of the total cross section from the threshold is dominantly
caused by the combination of the contributions from the $N(2120){3/2}^-$
resonance and the Born term. We mention that the contribution of the $K$ meson
exchange in our models is a little bit different from that in
Ref.~\cite{Kim:2014}, not only because the cutoff mass $\Lambda_K$ is fitted to
be around $1.0$ GeV in our model while it is fixed to be $1.1$ GeV in
Ref.~\cite{Kim:2014}, but also due to the pseudo-vector coupling chosen in our
models for the $\Lambda NK$ vertex (see Eq.~(\ref{eq:L_LNK})) in contrast to
the pseudo-scalar coupling adopted in Ref.~\cite{Kim:2014}.

We now come back to the issue of the dip structure exhibited by the
total cross section result of model IV which is caused by the interference of
the $N(2060){5/2}^+$ and $N(2190){7/2}^-$ resonances as can be seen from
Fig.~\ref{fig:total_cro_sec}. Clearly, the data do not show such a structure. A
careful inspection of the differential cross section fit results of model IV
reveals that this structure is due to the model not being able to quite describe the measured differential cross
section data at one energy, namely, at $W=2.217$ GeV. We note that, actually,
not only model IV, but all the other models are unable to quite reproduce the
angular behavior exhibited by the data at this energy. The
relevant difference between the results of model IV and the other models that
causes the valley structure in model IV is that model IV under-predicts the
data in the angular region of $0 \lesssim \cos\theta \lesssim 0.5$, while the
other models also under-predict in this angular region but over-predict for
other angles. As a result, the total cross section predictions for other models
agree with the data but it is under-predicted by model IV at this energy. In
fact, if one forces to describe the differential cross section better at
$W=2.217$ GeV (at the expenses of a slightly deteriorating description for the
neighboring energies), the dip structure in the prediction of model IV
disappears completely as illustrated in Fig.~\ref{fig:weight}.

As can be seen in Figs.~\ref{fig:fit23}-\ref{fig:fit36} and as has been discussed above, the models I-V describe the most recent CLAS data on the differential cross sections for $K^{*+}$ photoproduction fairly well overall and with similar fit qualities in the full energy-range considered. However, they exhibit quite different resonance contents, as indicated in Table~\ref{Table:para} and clearly seen in the total cross section predictions shown in Fig.~\ref{fig:total_cro_sec}. By now, it is a well known fact that the cross section data alone (even with high-statistics) do not impose enough stringent constraints on the fits to pin down the model parameters, especially, on the resonance contents and associated resonance parameters. Such a feature has also been found and discussed in Ref.~\cite{Nakayama:2006etap} in a study of $\eta'$ photoproduction. One expects that the spin observables may be more sensitive to the dynamical contents of different models. In Figs.~\ref{fig:beam_asy}-\ref{fig:recoil_asy}, we show the predictions of the photon beam asymmetry ($\Sigma$), target nucleon asymmetry ($T$), and recoil $\Lambda$ asymmetry ($P$) corresponding to our models I-V. As we can see, unlike the cross sections, the predictions for spin observables vary considerably among different models. For energies where the photon beam asymmetry is less sensitive to the models, the target nucleon asymmetry and the recoil $\Lambda$ asymmetry are quite sensitive, and {\it vice versa}. Therefore, overall, a combined analysis of the data of these spin observables is expected to impose much more stringent constraints on the resonance contents and help determine better the model parameters for $\gamma p \to K^{*+} \Lambda$. We hope that these spin observables can be measured in experiments in the near future.

\section{Summary and conclusion}  \label{sec:summary}

In the present work, we have analyzed the most recent high-statistic cross section data reported by the CLAS Collaboration for the $\gamma p \to K^{*+} \Lambda$ reaction \cite{Tang:2013}. The analysis has been based on an effective Lagrangian approach in the tree-level approximation. Apart from the $t$-channel $K$, $\kappa$, $K^*$ exchanges, the $s$-channel nucleon ($N$) exchange, the $u$-channel $\Lambda$, $\Sigma$, $\Sigma^*(1385)$ exchanges, and the generalized contact current, the contributions from the near-threshold nucleon resonances in the $s$-channel have been also taken into account in constructing the reaction amplitude. The generalized contact current introduced in the present work ensures that the  reaction amplitude in our model is fully gauge invariant as it obeys the generalized Ward-Takahashi identity \cite{Haberzettl:1997,Haberzettl:2006,Huang:2012,Huang:2013}.

It is found that to obtain a satisfactory description of the high-statistics differential cross section data from CLAS, at least two nucleon resonances should be included in the $s$-channel interaction diagrams. Furthermore, we have found five distinct sets of
resonances that describe these data with similar accuracies in the whole energy range of $1.75$ GeV $\leqslant E_\gamma \leqslant$ $3.85$ GeV. One of these two resonances, common to all five sets, is the $N(2060){5/2}^-$; the other resonance in each of the five sets is $N(2000){5/2}^+$, $N(2040){3/2}^+$, $N(2100){1/2}^+$, $N(2120){3/2}^-$ and $N(2190){7/2}^-$, respectively. The differential cross section data near the $\Lambda K^*$ threshold is for the first time described quite satisfactorily. The resulting resonance masses are compatible with those advocated by the Particle Data Group (PDG) \cite{Patrignani:2016}. Although the CLAS total cross section data -- which are obtained by integrating the measured differential cross sections -- may suffer from the limited angular acceptance of the CLAS detector for forward angles, the predicted total cross sections are in good agreement with these data.

It is shown that, together with the $K$-meson exchange, the $N(2060){5/2}^-$ resonance practically determine the dynamics of the $\gamma p \to K^{*+} \Lambda$ reaction in the low-energy region in the present model. In particular, they are responsible for the observed shape of the angular distribution and for the sharp raise of the total cross section from the threshold up to $W \sim 2.1$ GeV. The other resonance, in each of the five sets, contribute significantly at higher energies. The $K$ meson exchange provides a very significant contribution to the cross sections in the entire energy range considered, especially at high energies where it dominates the cross section to a large extent, while the contributions from the other non-resonant terms are found to be negligible.

The predicted photon beam asymmetry, target nucleon asymmetry, and recoil $\Lambda$ asymmetry are found to vary considerably from one model to another, indicating their sensitivity to the details of the models, in particular, to the different resonance contents that cannot be distinguished by the cross section alone. It is expected that the data for these spin observables would impose more stringent constraints on the models than the cross sections and help understand better the reaction mechanism and determine better the resonance contents and associated resonance parameters in the $\gamma p \to K^{*+}\Lambda$ reaction. We hope that these data can be measured in experiments in the near future.

We should also mention that although the present calculation describes the differential cross section data quite well overall and much better than any of the earlier calculations, the agreement with the data is not perfect. In particular, the details of the observed angular behavior at $W=2.217$ GeV is not quite described by any of our models I-V.  As mentioned in Sec.~\ref{Sec:results}, the inclusion of one more resonance doesn't help improve much the fit quality.  A further investigation is required here.

Finally, we mention that a more complete analysis for extracting nucleon resonances requires a coupled channels calculation. To our knowledge, such an approach, which is beyond the scope of the present work, has not yet been developed for $K^*$ meson production reactions.

\begin{acknowledgments}
The authors are grateful to I. Strakovsky and K. Hicks for their kind help with
the experimental data aspects, and thank S.-H. Kim for providing us with their
theoretical results. The work of F.H. is partially supported by the National Natural Science Foundation of China under
Grants No.~11475181 and No.~11635009, and the Youth Innovation Promotion
Association of CAS under Grant No.~2015358. W.L.W. is grateful to the support of the State Scholarship Fund provided by the China Scholarship Council under Grant No.~201506025055.
The work of H.H. is partially supported by the U.S. Department of Energy Award DE-SC0016582. The authors
acknowledge the National Supercomputing Center of Tianjin in China and the
J\"{u}lich Supercomputing Center at Forschungszentrum J\"{u}lich in Germany for
providing computing resources that have contributed to the research results
reported within this paper. 
\end{acknowledgments}

\end{document}